\documentclass[a4paper,10pt]{article}

\usepackage{amssymb, amsthm}
\usepackage{amsmath}
\usepackage[usenames,dvipsnames]{color}
\usepackage[colorlinks, citecolor=blue, linkcolor=blue, urlcolor=Maroon, filecolor=Maroon]{hyperref}
\usepackage{epsfig,psfrag}

\newtheorem{theo}{Theorem}[section]
\newtheorem{prop}[theo]{Proposition}
\newtheorem{coroll}[theo]{Corollary}

\theoremstyle{definition}

\newcommand{\sfrac}[2]{{\textstyle\frac{#1}{#2}}}
\newcommand{\und}{\qquad\textrm{and}\qquad}
\newcommand{\RR}{{\mathbb{R}}}

\newcommand{\ii}{{\rm i}}
\newcommand{\dd}{{\rm d}}

\newcommand{\Tr}{\mathrm{Tr}}

\title{Instantons and Killing spinors}
\author{Derek Harland${}^\ast$ and Christoph N\"olle${}^\dagger$
\bigskip \\
\noindent ${}^\ast$\normalsize{Department of Mathematical Sciences, Durham University,}\\
\normalsize{Science Laboratories, South Road, Durham, DH1 3LE, UK}\\
\normalsize{\texttt{d.g.harland@durham.ac.uk}}
\bigskip \\
\noindent ${}^\dagger$\normalsize{Institut f\"ur Theoretische Physik, Leibniz Universit\"at Hannover,}\\
\normalsize{Appelstra{\ss}e 2, 30167 Hannover, Germany}\\
\normalsize{\texttt{noelle@math.uni-hannover.de}}
}
\date{19th September 2011}

\begin{document}
\maketitle
\numberwithin{equation}{section}

\parindent=0cm
\parskip=\bigskipamount

\begin{abstract}
We investigate instantons on manifolds with Killing spinors and their cones. Examples of manifolds with Killing spinors
include nearly K\"ahler 6-manifolds, nearly parallel $G_2$-manifolds in dimension 7, Sasaki-Einstein manifolds, and
3-Sasakian manifolds.  We construct a connection on the tangent bundle over these manifolds which solves the instanton
equation, and also show that the instanton equation implies the Yang-Mills equation, despite the presence of torsion. 
We then construct instantons on the cones over these manifolds, and lift them to solutions of heterotic supergravity. 
Amongst our solutions are new instantons on even-dimensional Euclidean spaces, as well as the well-known BPST,
quaternionic and octonionic instantons.
\end{abstract}

\section{Introduction and summary}

Manifolds with real Killing spinors frequently occur as supersymmetric backgrounds in string theory \cite{Acharya98,
Koerber08}. Such manifolds are Einstein, and they always admit a $G$-structure, that is, a reduction of the structure
group of their tangent bundle from $\mbox{SO}(n)$ to $G$, where $G$ is some Lie subgroup of $\mbox{SO}(n)$.  This Lie
group $G$ is not however the holonomy group of the Levi-Civita connection, so the $G$-structure is not integrable. 
Nevertheless, manifolds with real Killing spinors have a close kinship with manifolds with special holonomy: the cone
metric over a manifold with real Killing spinor does have special holonomy.  This observation allowed B\"ar to classify
manifolds with real Killing spinors \cite{Baer93}.  Besides the round spheres, the only manifolds with real Killing
spinors are nearly parallel $G_2$-manifolds in dimension 7, nearly K\"ahler manifolds in dimension 6, Sasaki-Einstein
manifolds, and 3-Sasakian manifolds.

Instanton equations in dimensions greater than 4 were first written down almost 30 years ago \cite{CDFN83, Ward84}.  It was
later realised that many of these equations are naturally BPS, so play a role in supersymmetric theories, including
heterotic supergravity.  The instanton equations make sense on any manifold with a $G$-structure,
and it is hoped that their study will result in new invariants for such manifolds, just as the original instanton
equations were the main ingredient in Donaldon's 4-manifold invariants \cite{DT98,DS09,Tian00}.  Thus the search for solutions
to the instanton equations is well-motivated, and many examples of instantons have appeared in the literature
\cite{FN84,FN85,CGK85,IP92,BIL08,HILP10,HP10,HILP11,Gemmer11,Xu09,Correia09,Correia10}.

On manifolds with integrable $G$-structures instanton equations have the following two important features: they imply
the Yang-Mills equation; and they have a distinguished solution on the tangent bundle, namely the Levi-Civita
connection.  On manifolds with non-integrable $G$-structure neither of these properties is expected to hold true in
general.  The first purpose of the present article is to show that both properties do hold on manifolds with real
Killing spinors.  In doing so we construct a distinguished connection on the tangent bundle which solves the instanton
equation, and which seems to be an analog of the Levi-Civita connection in the geometry of real Killing spinors.

The second purpose of this article is to construct solutions of the instanton equation on the cone over a manifold with
real Killing spinor, and to lift them to solutions of the BPS equations and Bianchi identity of heterotic
supergravity.  We find a 1-parameter family of instantons on the cone over any manifold with real Killing spinor.  Our construction proceeds by making an ansatz which reduces the instanton equations to ODEs; remarkably,
this procedure works without assuming that the underlying manifold has any symmetries, so seems to be an example of a
``consistent reduction'' \cite{GV07}.  Our construction of instantons on cones generalises one
given in the Sasaki-Einstein case in \cite{Correia09}, and the lift to supergravity generalises the well-known
constructions \cite{Strom90,HS90,GN95}. 

Our construction can in particular be applied to cones over spheres.  Doing so reproduces many known instantons on Euclidean
spaces, including the BPST instanton on $\RR^4$ \cite{BPST75}, the octonionic instantons on $\RR^7$ and $\RR^8$
\cite{FN84,FN85,IP92,GN95}, and the quaternionic instantons on $\RR^{4m+4}$ \cite{CGK85,BIL08}, and also produces a new family of
hermitian instantons on even-dimensional Euclidean spaces.  All of these instantons come equipped with a size parameter.  In the limit of zero size one obtains instantons with point-like singularities.  Thus our instantons on Euclidean spaces provide simple examples of singularity-formation: the limiting singular connections are examples of Tian's ``tangent connections''\cite{Tian00}.

It happens that the cones over many known manifolds with real Killing spinors admit smooth resolutions, so an obvious
next step is to consider instantons on these resolutions -- in fact, this has already been done in the Sasaki-Einstein
case \cite{Correia10}.  We hope to report on this in the future.

 One particular motivation to look for string solitons on cones over Killing spinor manifolds was the discovery of 
heterotic supergravity backgrounds with linear dilaton on the cylinder over certain non-symmetric homogeneous spaces in
\cite{Noe10}. In
4 dimensions such solutions occur as the near horizon limit of NS5-branes \cite{CHS91a}; the full supergravity brane
solution interpolates between $\mathbb R\times S^3$ with linear dilaton and flat $\mathbb R^4$. It has enhanced
supersymmetry as compared to the similar solutions on $\mathbb R^4$ found by Strominger in \cite{Strom90}, and does not
receive any $\alpha'$-corrections. The lecture notes \cite{CHS91b} contain a review of the results of \cite{Strom90} and
\cite{CHS91a}. The solutions to be presented here do not generalize the NS5-branes of \cite{CHS91a}, but
instead the results of \cite{Strom90}. In particular, the linear
dilaton solutions of \cite{Noe10} do not appear as a limiting case of our backgrounds.

This article is arranged as follows.  In section \ref{sec:inst} we discuss various formulations of the instanton
equations, and show that they imply the Yang-Mills equation on manifolds with real Killing spinors.  For completeness we
also give the spinorial viewpoint on the Hermitian-Yang-Mills equations.  In section
\ref{sec:geom} we describe in detail the geometry of manifolds with real Killing spinors, and construct the connections
on the tangent bundles of these manifolds which solve the instanton equations.  In section \ref{sec:cone} we construct
instantons on the cones over these manifolds, and in section \ref{sec:string} we lift these to solutions of heterotic
supergravity.

\paragraph{Conventions.} Before beginning we outline our conventions.  We will always work with an orthonormal frame
$e^\mu$ for the cotangent bundle, where $\mu,\nu,\ldots$ are generic indices; the dual frame of vector fields will be
denoted $L_\mu$.  We will adopt the shorthand $e^{\mu\nu}=e^\mu\wedge e^\nu$ etc.  Forms $\theta$ map to elements of the
Clifford algebra using the standard map
\begin{equation}
 \sfrac{1}{p!}\theta_{\mu_1\cdots\mu_p} e^{\mu_1\cdots \mu_p} \mapsto \sfrac{1}{p!}\theta_{\mu_1\cdots\mu_p}
\gamma^{\mu_1}\cdots \gamma^{\mu_p} = \sfrac{1}{p!}\theta_{\mu_1\cdots\mu_p} \gamma^{\mu_1\cdots \mu_p}.
\end{equation}
Here $\gamma^\mu$ are Clifford matrices satisfying $\{\gamma^\mu,\gamma^\nu\}=2g^{\mu\nu}=2\delta^{\mu\nu}$, and
$\gamma^{\mu_1\cdots\mu_p}$ denotes a totally anti-symmetrised Clifford product. The Clifford action of a form
$\theta$ on a spinor $\epsilon$ is denoted by $
   \theta \cdot \epsilon.$
Connections on the tangent bundle will
be represented by matrix-valued 1-forms $\Gamma_\mu^\nu=e^\kappa\Gamma_{\kappa\mu}^\nu$, so that the covariant
derivative of a 1-form $v=v_\mu e^\mu$ is $\nabla v = (\dd v_\mu - v_\nu\Gamma_\mu^\nu)\otimes e^\mu$, and the covariant
derivative of a spinor $\psi$ is $\nabla\psi=\dd\psi + \sfrac14\Gamma^\mu_\nu\gamma_\mu\gamma^\nu\psi$.  The torsion $T^\mu$ of a connection $\Gamma_\mu^\nu$ can be calculated using the Cartan structure equation:
\begin{equation}
\label{CSE}
 T^\mu = \dd e^\mu + \Gamma^\mu_\nu\wedge e^\nu.
\end{equation}
Indices $\alpha,\beta,\ldots$ will run from 1 to 3, and indices $a,b,\ldots$ will have specific ranges, to be explained in
section \ref{sec:geom}.

\section{Instantons and the Yang-Mills equation}
\label{sec:inst}

Let $E\rightarrow M$ be a vector bundle over a Riemannian manifold $(M,g)$ of dimension $n$, and $A$ a connection on $E$
with
curvature
\begin{equation}
 F=dA+A\wedge  A \ \in\ \Gamma(\Lambda^2 T^*M \otimes\, \text{End}(E)).
\end{equation}
There are many different ways to define an instanton condition for $F$.  The first way, which will be central to this
paper, is valid when $M$ is a spin manifold, and the spinor bundle admits one or more non-vanishing spinors $\epsilon$. 
Then $A$ will be called an instanton if
\begin{equation}
\label{instcond1}
 F\cdot \epsilon = 0.
\end{equation}
This instanton condition is natural in supersymmetric theories, where the spinor $\epsilon$ can be identified with a
generator of supersymmetries.

The second definition of an instanton is valid when $(M,g)$ is equipped with a $G$-structure, that is, a reduction of
the structure group of the tangent bundle to a Lie subgroup $G\subset\mbox{SO}(n)$.  This means that at every point in
$M$ there exists a Lie-subalgebra $\mathfrak{g}\subset\mathfrak{so}(n)$ which acts on tangent vectors.  This can be
identified with a subspace $\mathfrak{g}\subset\Lambda^2(\RR^n)$, using the canonical isormorphism
$\mathfrak{so}(n)\cong\Lambda^2(\RR^n)$ induced by the metric; then $A$ is called an instanton if the 2-form part of $F$
belongs to this subspace.  In global terms, $F$ is an instanton if
\begin{equation}
\label{instcond2}
 F \in \Gamma(W \otimes\, \text{End}(E)) \subset \Gamma(\Lambda^2 T^*M \otimes\, \text{End}(E)),
\end{equation}
where $W\subset\Lambda^2T^*M$ is the vector bundle with fibre $\mathfrak{g}$.  This condition is often abreviated
to $F\in\mathfrak{g}$, and we will do so here.

The third definition of instanton also exploits a $G$-structure.  If $\mathfrak{g}$ is simple, then its quadratic
Casimir is an element of $\mathfrak{g}\otimes\mathfrak{g}$ invariant under the action of $G$, which may be identified
with a section of $\Lambda^2\otimes\Lambda^2$ and hence is mapped to a section $Q$ of $\Lambda^4$ by taking a wedge
product. It turns out that $Q$ vanishes for SO($n$), but is non-trivial for any other simple Lie group. Since $Q$ is by
construction $G$-invariant, the operator $u\mapsto\ast(\ast Q\wedge u)$ acting on 2-forms $u$ commutes with the action
of $G$, so by Schur's lemma the irreducible representations of $G$ in $\Lambda^2$ are eigenspaces for $Q$. Then $A$ is
called an instanton if $F$ belongs to one of these eigenspaces, that is, if
\begin{equation}
\label{instcond3}
 \ast Q\wedge F = \nu\ast F
\end{equation}
for some $\nu\in\RR$.

These three definitions of instanton are related to each other.  The first definition is a special case of the second,
where $G\subset\mbox{SO}(n)$ is a subgroup which fixes the spinor(s) $\epsilon$ (assuming that this subgroup is the same
at all points of $M$).  And the second definition is a special case of the third, as long as $G$ is simple, since the
subspace $\mathfrak{g}\subset\Lambda^2$ forms an irreducible sub-representation.  The third definition was introduced in
\cite{CDFN83}, and predates the others.  In the case $n=4$ when $\epsilon$ is a Weyl spinor with positive helicity the first and second definitions are equivalent to the
anti-self-dual equation, while the third definition includes both the anti-self-dual and self-dual equations.

In this paper we will be interested only in the first definition of an instanton, but the second and third will prove
useful in calculations.  For the most part, we will specialise to the case where $\epsilon$ satisfy the equation
\begin{equation}
 \label{Killingeq}
 \nabla^{LC}_\mu \epsilon = \ii\lambda\gamma_\mu\cdot\epsilon,
\end{equation}
where $\nabla^{LC}$ is the Levi-Civita connection, $\gamma_\mu$ are a representation of the Clifford algebra, and
$\lambda$ is a real constant.  If $\lambda=0$ then $\epsilon$ are parallel and $(M,g)$ is obviously a manifold of
special holonomy.  If $\lambda\neq0$ then the $\epsilon$ are called real Killing spinors, and by rescaling the metric
and adjusting orientations, one can always arrange that $\lambda=1/2$.  The cone over $M$ is the manifold $\RR\times M$
equipped with metric
\begin{eqnarray}
 g_C = e^{2\tau}( \dd\tau^2 + g) = \dd r^2 + r^2 g,
\end{eqnarray}
where $\tau\in\RR$ and $r=e^\tau$.  It was first noticed by B\"ar that Killing spinors on $M$ lift to parallel spinors
on the cone, and this lead to a classification of manifolds with real Killing spinor \cite{Baer93}.

Instanton equations were originally introduced as a means of solving the Yang-Mills equation.  The traditional way of
relating the instanton equation to the Yang-Mills equation utilises the third definition.  By applying the exterior
derivative to \eqref{instcond3} and using the Bianchi identity, one obtains for $\nu\neq0$
\begin{equation}
 \label{YMeq}
 \nabla^A\wedge\ast F -\frac{1}{\nu} \dd\ast Q\wedge F = 0,
\end{equation}
where $\nabla^A\wedge\ast F$ is shorthand for $\dd\ast F +A\wedge\ast F +(-1)^{n-1}\ast F\wedge A$.  On manifolds of
special
holonomy the 4-form $Q$ is both closed and coclosed, so the second term vanishes and we are left with the Yang-Mills
equation $\nabla^A\wedge\ast F=0$. 

If $M$ is a manifold with real Killing spinor, $Q$ is not coclosed, so a priori the second term does not vanish. 
Nevertheless, the instanton equation does imply the Yang-Mills equation on a manifold with real Killing spinor, as the
following proposition shows:
\begin{prop}\label{prop_YM+Spinors}
 Suppose that $M$ is spin and carries a spinor $\epsilon$ solving equation \eqref{Killingeq}.  If $A$ is gauge field on
$M$ whose curvature form satisfies equation \eqref{instcond1}, then it solves the Yang-Mills equation.
\end{prop}
\begin{proof}
 The main idea of the proof is to act on equation \eqref{instcond1} with a Dirac operator constructed from the
Levi-Civita connection and $A$:
 \begin{equation}
  \mathcal{D} = \gamma^\mu\mathcal{D}_\mu = \gamma^\mu\left(L_\mu +
\frac{1}{4}\Gamma_{\mu\kappa}^\nu\gamma_\nu\gamma^\kappa + A_\mu\right)
 \end{equation}
 The Levi-Civita connection defines a covariant derivative on 2-forms,
 \begin{equation}
  (\nabla^{LC}_\mu F)_{\nu\kappa} = L_\mu F_{\nu\kappa} - \Gamma_{\mu\nu}^\lambda
F_{\lambda\kappa}-\Gamma_{\mu\kappa}^\lambda F_{\nu\lambda},
 \end{equation}
 and this satisfies the identities,
 \begin{eqnarray}
  \frac{1}{2}e^{\mu\nu\kappa}(\nabla^{LC}_\mu F)_{\nu\kappa} &=& \dd F \\
  g^{\mu\nu}e^{\kappa}(\nabla^{LC}_\mu F)_{\nu\kappa} &=& (-1)^n\ast\,\dd\ast F.
 \end{eqnarray}
 It follows that
 \begin{eqnarray}
  \gamma^\mu [\mathcal{D}_\mu ,F] &=& \frac{1}{2}\gamma^\mu \gamma^{\nu\kappa} ( (\nabla^{LC}_\mu F)_{\nu\kappa} +
[A_\mu ,F_{\nu\kappa}]) \\
  &=& \frac{1}{2}(\gamma^{\mu\nu\kappa}+g^{\mu\nu}\gamma^\kappa-g^{\mu\kappa}\gamma^\nu)( (\nabla^{LC}_\mu
F)_{\nu\kappa} + [A_\mu ,F_{\nu\kappa}]) \\
  &=& \nabla^A\wedge F + (-1)^n\ast (\nabla^A\wedge\ast F) \\
  &=& (-1)^n\ast (\nabla^A\wedge\ast F).
 \end{eqnarray}
 Therefore acting on \eqref{instcond1} with the Dirac operator gives
 \begin{eqnarray}
  0 &=& \mathcal{D}(F\cdot\epsilon) \\
  &=& (-1)^n\ast (\nabla^A\wedge\ast F)\cdot\epsilon + \gamma^\mu F \cdot\nabla^{LC}_\mu \epsilon.
 \end{eqnarray}
 Thus far we have not employed the spinor equation \eqref{Killingeq}.  This equation, together with the identity
$\gamma^\mu F \gamma_\mu=(n-4)F$ implies that
 \begin{eqnarray}
  0 = (-1)^n\ast (\nabla^A\wedge\ast F)\cdot\epsilon + \ii\lambda(n-4)F\cdot\epsilon.
 \end{eqnarray}
 The instanton equation \eqref{instcond1} implies that the second term vanishes, and since the action of 1-forms on
spinors is invertible, we conclude that $\nabla^A\wedge\ast F=0$.
\end{proof}
Note that this proposition applies to manifolds with parallel spinor as well as manifolds with real Killing spinor.  The
special case of this proposition where $M$ is nearly K\"ahler was previously obtained using a different method by Xu
\cite{Xu09}.  We will give some alternative proofs of this proposition in the following section.

The existence of globally defined spinors seems to be essential for instantons to satisfy the
Yang-Mills equation. For instance, on K\"ahler manifolds with holonomy group U($m)$ the most obvious
instanton condition $F\in \mathfrak u(m)$ does not automatically imply the Yang-Mills equation, because U$(m$) does
not fix any spinor.

Thus in order to obtain solutions of the Yang-Mills equations on K\"ahler manifolds, a stronger instanton equation is needed.
The holonomy Lie algebra splits as
$\mathfrak{u}(m)=\mathfrak{su}(m)\oplus\mathfrak{u}(1)$, so there exist subspaces
$\mathfrak{su}(m),\mathfrak{u}(1)\subset\Lambda^2$ --
where $\mathfrak{u}(1)$ is just the subspace spanned by the K\"ahler form $\omega$. One possibility is to impose the
stronger equation
$F\in \mathfrak{su}(m)$, but this excludes many interesting examples, such as the Levi-Civita connection on a Hermitian
symmetric space. To cover this case as well, but without losing the Yang-Mills equation, the instanton condition
on K\"ahler manifolds involves the requirement $F\in \mathfrak{u}(m)$ and an additional
constraint on the $\mathfrak u(1)$-part of $F$, known as the Hermitian-Yang-Mills equation \cite{DUY},
 \begin{equation}\label{HYM}
   F^{\mathfrak u(1)} = \mu\,\omega \otimes  J,
 \end{equation}  
 where $\mu \in \mathbb R$ and $J\in $ End$(E)$ is a constant central element in the Lie algebra of the gauge group.
 
 The Hermitian-Yang-Mills equation implies the Yang-Mills equation, and this can be proven by spinorial methods as well. Although a general K\"ahler manifold does not possess a parallel spinor or even a spin bundle, the tensor product of the spin bundle with a square root of the canonical bundle is well-defined and has a parallel section $\epsilon'$. Equation \eqref{HYM} implies that $F^{\mathfrak u(1)}$ and $F^{\mathfrak {su}(m)}$ satisfy the separate Bianchi identities
 \begin{equation}
   \dd F^{\mathfrak u(1)} = \nabla^A \wedge F^{\mathfrak {su}(m)}=0,
 \end{equation} 
 and the proof of proposition \ref{prop_YM+Spinors} goes through for $F^{\mathfrak {su}(m)}$ and $\epsilon'$ instead of $F$ and $\epsilon$. Hence the two components of $F$ satisfy two separate Yang-Mills equations
 \begin{equation}
   \dd*F^{\mathfrak u(1)}  = \nabla^A \wedge * F^{\mathfrak {su}(m)}=0,
 \end{equation}  
 which imply in particular the full Yang-Mills equation for $F$.

\section{Geometry of real Killing spinors}
\label{sec:geom}

 From this section on, our attention will be focused on manifolds $M$ with real Killing spinor.  Specifically, $M$ will
be either 7d nearly parallel $G_2$, 6d nearly K\"ahler, $(2m+1)$-dimensional Sasaki-Einstein or $(4m+3)$-dimensional 3-Sasakian
(so we are neglecting even-dimensional spheres in dimensions other than 6).  The Killing spinors $\epsilon$ define a
$K$-structure, where $K=G_2$, SU(3), $\mbox{SU}(m)$ or $\mbox{Sp}(m)$ respectively.
 
 These manifolds share a number of common properties.  They all come equipped with a canonical 4-form $Q'$ and 3-form
$P'$, defined by
 \begin{equation}
  \begin{aligned}
   P' &= -\frac{\ii}{3!}\big\langle \epsilon , \gamma_{\mu\nu\rho} \epsilon \big\rangle e^{\mu\nu\rho} \\
   Q' &= -\frac{1}{4!}\big\langle \epsilon , \gamma_{\mu\nu\kappa\lambda} \epsilon \big\rangle e^{\mu\nu\kappa\lambda},
  \end{aligned}
 \end{equation}
 and we normalise them by fixing $\langle\epsilon,\epsilon\rangle=1$.  Since these forms are constructed as bilinears in
the Killing spinors, they are parallel with respect to any connection with holonomy group $K$.  The Killing spinor equation
implies that these satisfy the differential identities,
 \begin{equation}
 \label{dPdQ}
  \dd P'=4Q',\quad \dd \ast Q' = (n-3)\ast P'. 
 \end{equation}
 It follows that the 4-form
 \begin{equation}
  e^{4\tau}(\dd\tau\wedge P' + Q')
 \end{equation}
 on the cone $\RR\times M$ is both closed and co-closed -- in fact, this is the Casimir 4-form associated to the
$G$-structure on the cone.
 
 Associated to the $K$-structure on $M$ is a Casimir 4-form $Q$, which we normalise so that the instanton equation
\eqref{instcond1} is equivalent to
 \begin{equation}
  \ast F=-\ast Q\wedge F.
 \end{equation}
 It turns out that $Q$ is always exact on real Killing spinor manifolds, so that one can also find a 3-form $P$
which satisfies $\dd P=4Q$ and which is parallel with respect to any connection with
holonomy $K$.  On the nearly parallel $G_2$, nearly K\"ahler, and Sasaki-Einstein manifolds $Q=Q'$ and $P=P'$, but on
3-Sasakian manifolds this is not the case.
 
 We will call a connection on the tangent bundle of a manifold with $K$-structure \emph{canonical} if it has holonomy
$K$ and torsion totally antisymmetric with respect to some $K$-compatible metric.  All of the real Killing spinor manifolds that we consider come equipped
with a canonical connection $\nabla^P$ on the tangent bundle, which we construct on a case-by-case basis.  In all cases
the torsion is proportional to the parallel 3-form $P$.  The significance of the canonical connection is that it is an
instanton.  This follows from a general proposition \ref{prop:TorsionExSymmetry} which we state and prove at the end of this section.  Our canonical connection differs in
subtle ways from the characteristic connection introduced in \cite{FI01,Agricola06}, and we will clarify exactly how at
the end of this section.  Also in this section we will supply some alternative proofs of proposition \ref{prop_YM+Spinors}.
 
 A key idea that will be used in this section and throughout this article is the relation between parallel objects and
trivial representations, sometimes known as the general holonomy principle \cite{Agricola06}.  Suppose that $B\to M$ is
a principal bundle with structure group $K$, and let $V$ be a vector space which forms a representation of $K$.  Then
there is an associated vector bundle with fibre $V$.  Any $K$-invariant vector $v\in V$ lifts to a global non-vanishing
section of the bundle, and this section will be parallel with respect to any $K$-connection.  Thus, the study of
parallel objects on a vector bundle reduces to linear algebra; in particular, this procedure allows us to construct
parallel forms and spinors.

\subsection{Nearly parallel $\mathbf{G_2}$}
 The stabiliser of a Majorana spinor in 7 dimensions is the exceptional group $G_2$.  Thus a 7-manifold with 1 real
Majorana Killing spinor admits a $G_2$-structure.  The canonical 3- and 4-forms $P=P'$, $Q=Q'$ satisfy $P=\ast Q$, and one
can choose a local orthonormal frame $e^a$, $a=1,\dots,7$ so that they take the standard forms,
 \begin{equation}
  \begin{aligned}
   P &= e^{123} + e^{145} - e^{167} + e^{246} + e^{257} + e^{347} - e^{356} \\
   Q &= e^{4567} +e^{2367} -e^{2345} +e^{1357} +e^{1346} +e^{1256} -e^{1247}.
  \end{aligned}
 \end{equation}
The fact that $\dd P=4Q$, then implies that the $G_2$-structure is nearly parallel.

\paragraph{Canonical connection.}
 The canonical connection is constructed by perturbing the Levi-Civita connection by the 3-form $P$.  The 3-form $P$
acts with eigenvalue $7\ii$ on $\epsilon$, and with eigenvalue $-\ii$ on the 7-dimensional orthogonal complement of
$\epsilon$.  For any $\gamma_a$, $\gamma_a\cdot\epsilon$ is orthogonal to $\epsilon$.  It follows that
 \begin{equation}
  P_{abc}\gamma^{bc} \epsilon = (\gamma_a\cdot P + P\cdot\gamma_a)\epsilon = 6\ii\gamma_a\cdot\epsilon
 \end{equation}
 (the same result can also be obtained using Fierz identities).  We define the canonical connection by the equation
 \begin{equation}
  {}^P\Gamma_{ab}^c = {}^{LC}\Gamma_{ab}^c +\frac{1}{3} P_{abc}.
 \end{equation}Then it follows from the Killing spinor equation \eqref{Killingeq} and the identity proved above that
 \begin{equation}
  \nabla^P\epsilon = 0.
 \end{equation}
 Therefore $\nabla^P$ has holonomy $G_2$.  The torsion of $\nabla^P$ can be calculated from the Cartan structure
equation \eqref{CSE}, and is
 \begin{equation}
 \label{NP torsion}
  T^a = \frac{1}{3}P_{abc}e^{bc}.
 \end{equation}
 We note that although we have only defined the canonical connection using a local frame, it is nonetheless globally
well-defined, because it is constructed from the Levi-Civita connection and the global 3-form $P$.
 
\paragraph{$G_2$-instantons.}
 The instanton equation \eqref{instcond1} is equivalent to \eqref{instcond3} with $\nu=-1$:
 \begin{equation}\label{G2instEq1}
   *F = - \ast Q \wedge F = -P \wedge F
 \end{equation}  
 The two-forms decompose as $\Lambda^2 \simeq \underbar{14} \oplus \underbar {7}$ under $G_2$, where $\underbar{14}$ is
the adjoint and $\underbar 7$ the fundamental representation.  As explained in the introduction, the instanton equation
is equivalent to $F$ being in the adjoint representation \eqref{instcond2}.  Now $Q$ is $G_2$-invariant, so that $Q
\wedge F \in \Lambda^6 $ must be in the same representation as $F$, but $\Lambda^6 \simeq \underbar 7$ is actually the
fundamental representation. It follows that \eqref{G2instEq1} is equivalent to
\begin{equation}\label{G2instEq2}
 F\wedge Q =0.
\end{equation}  
Applying the Yang-Mills operator to \eqref{G2instEq1} leads to 
 \begin{equation}
   \nabla^A\wedge * F + 4Q \wedge F=0,
 \end{equation} 
 but the torsion term $Q\wedge F$ vanishes due to \eqref{G2instEq2}.  Thus the instanton equation implies the Yang-Mills
equation, confirming proposition \ref{prop_YM+Spinors}.
 
\paragraph{Examples.}
 Simply connected nearly parallel $G_2$-manifolds with two Killing spinors are Sasaki-Einstein, and those with three
Killing spinors are 3-Sasakian. More than three Killing spinors exist only on the round sphere $S^7$. The            
following examples with exactly one Killing spinor are known \cite{Friedr97}. First of all, the Aloff-Wallach spaces
$N(k,l)=$ SU(3)/U(1)$_{k,l}$, where U(1) embeds
into SU(3) as
 \begin{equation}
     z \mapsto \text{diag}(z^k,z^l,z^{-(k+l)})
 \end{equation} 
 for $z\in S^1$ and positive integers $k,l$, each carry two homogeneous metrics with at least one Killing spinor. For
$(k,l)\neq (1,1)$ they both have exactly one Killing spinor, whereas for $(k,l)=(1,1)$ one of the two metrics is
3-Sasakian. Another homogeneous example is the Berger space SO(5)/SO(3)$_{\text{max}}$, where SO(3) acts on the tangent space by its unique irreducible 7-dimensional representation.
Additionally, every 3-Sasakian manifold in dimension 7 has a second Einstein metric with exactly one Killing spinor,
which gives some further examples. This metric will be described in paragraph \ref{ssec:3Sgeom} below. In particular, this construction gives rise to an additional nearly parallel $G_2$-structure on $S^7$, the so called squashed seven-sphere.

 The Aloff-Wallach spaces, Berger space and the squashed 7-sphere all have positive sectional curvature \cite{Ziller07, VZiller07}, and this seems to be true for many of the nearly parallel $G_2$ metrics obtained from 3-Sasakian manifolds as well \cite{Dearr04, BGbook}.

\subsection{Nearly K\"ahler}
 A Majorana spinor in 6 dimensions is fixed by the subgroup $\mbox{SU}(3)\subset\mbox{SO}(6)$, so a 6-manifold with a
Majorana Killing spinor has an SU(3)-structure.  In addition to the canonical 4- and 3-forms $Q=Q'$, $P=P'$ there are
parallel 2- and 3-forms $\ast Q=\omega$, $\ast P$.  One can choose a local orthonormal frame $e^a$, $a=1,\dots 6$ so
that these parallel forms take the standard form
 \begin{equation}
  \begin{aligned}
  \omega = e^{12}+e^{34}+e^{56},\quad Q=e^{1234}+e^{1256}+e^{3456},\\
  P+\ii\ast P = (e^1+\ii e^2)\wedge(e^3+\ii e^4)\wedge (e^5+\ii e^6).
  \end{aligned}
 \end{equation}
 Since $\dd P= 4Q$ and $\dd\ast Q=3\ast P$, the SU(3)-structure is nearly K\"ahler.

\paragraph{Canonical connection.} 
 Again, the canonical connection is constructed by perturbing the Levi-Civita connection by the 3-form $P$.  It can be
shown that
 \begin{equation}
  P_{abc}\gamma^{bc}\cdot \epsilon =  4\ii\gamma_a\cdot\epsilon.
 \end{equation}
 We define the canonical connection by
 \begin{equation}
  {}^P\Gamma_{ab}^c = {}^{LC}\Gamma_{ab}^c + \frac{1}{2}P_{abc}.
 \end{equation}
 Then it follows from the Killing spinor equation \eqref{Killingeq} and the identity proved above that
 \begin{equation}
  \nabla^P\epsilon = 0.
 \end{equation}
 Therefore $\nabla^P$ has holonomy SU(3).  The subgroup $\mbox{SU}(3)\subset\mbox{SO}(6)$ actually fixes two spinors, so
$\nabla^P$ has two parallel spinors; the second is obtained by acting on $\epsilon$ with the chirality operator. It is
a Killing spinor as well, but with opposite sign of the the Killing constant $\lambda$. The torsion of $\nabla^P$ can be
calculated from the Cartan structure equation \eqref{CSE}, and is
 \begin{equation}
 \label{NK torsion}
  T^a = \frac{1}{2}P_{abc}e^{bc}.
 \end{equation}
 
\paragraph{SU(3)-instantons.}
 The SU(3)-instanton equation \eqref{instcond1} is equivalent to
\begin{equation}\label{nKInstEq1}
 *F= - \omega\wedge F,
\end{equation} 
 and also to
 \begin{equation}
   F\in \Omega^{(1,1)}\und \langle \omega,F\rangle = 0.
 \end{equation} 
 In the form \eqref{nKInstEq1} it implies the Yang-Mills equation with torsion 
 \begin{equation}
   \nabla^A \wedge * F + 3 F\wedge \ast P=0,
 \end{equation} 
 but the torsion term vanishes due to $F\in \Omega^{(1,1)}$ and $\ast P\in \Omega^{(3,0)}\oplus
\Omega^{(0,3)}$. Thus $F$ satisfies the ordinary Yang-Mills equation, confirming proposition \ref{prop_YM+Spinors}. This
argument is due to Xu \cite{Xu09}.

\paragraph{Examples.}
There are precisely 4 homogeneous nearly K\"ahler manifolds, and they are $S^6=G_2/$SU(3),
$S^3\times S^3=$ SU(2)$^3$/SU(2)$_{\text{diag}}$, SU(3)/U(1)$^2$, and Sp(2)/Sp(1)$\times $U(1) \cite{But}.  Currently,
complete non-homogeneous examples are not known, but there exists a nearly K\"ahler structure with two conical
singularities on the so-called sine-cone over every 5-dimensional Sasaki-Einstein manifold, giving rise to incomplete
non-homogeneous examples \cite{FIMU06}.

\subsection{Sasaki-Einstein}

The subgroup $\mbox{SU}(m)\subset\mbox{Spin}(2m+1)$ fixes a 2-dimensional space of Dirac spinors.  These two spinors
transform with weights $\pm 1$ under the action of the centraliser U(1) of $\mbox{SU}(m)$, and will be labelled
$\epsilon,\tilde\epsilon$.  A $(2m+1)$-dimensional Sasaki-Einstein manifold can be defined to be a Riemannian manifold
with two Killing spinors $\epsilon,\tilde \epsilon$, so in particular admits an $\mbox{SU}(m)$-structure. The Killing
constants of the two spinors coincide for odd $m$, but have opposite sign for even $m$, as we shall see below. We
assume that $\epsilon$ satisfies the Killing spinor equation \eqref{Killingeq} with constant $\lambda = 1/2$.
 
From the Killing spinor $\epsilon$ one can construct parallel forms, this time with arbitrary degree.  Besides $P=P'$,
$Q=Q'$, we will need only the first two:
\begin{equation}
\label{SE bilinears}
 \begin{aligned}
  \eta &= \big\langle \epsilon , \gamma_\mu \epsilon \big\rangle e^\mu \\
  \omega &= -\sfrac{\ii}{2}\big\langle \epsilon , \gamma_{\mu\nu} \epsilon \big\rangle e^{\mu\nu}.
 \end{aligned}
\end{equation}
These forms are related to one another as follows:
\begin{equation}
 P = \eta\wedge\omega,\quad Q = \frac12\omega\wedge\omega,\quad \eta\lrcorner\,\omega=0.
\end{equation}
It will be convenient to pick an orthonormal basis $e^1,e^a$ with $e^1=\eta$ and $a=2,\dots,2m+1$, so that
\begin{equation}
 \eta=e^1,\quad \omega = e^{23}+e^{45}+\dots +e^{2m\,2m+1}.
\end{equation}
The Killing spinor equation implies that $\dd\eta=2\omega$ and $\dd\ast\omega=2m\ast\eta$, as well as $\dd P=4Q$ and
$\dd \ast Q = (2m-2)\ast P$.  For an extensive review of Sasakian geometry, we recommend the book \cite{BGbook}. A more
condensed review of Sasaki-Einstein manifolds can be found in \cite{Sparks10}.

\paragraph{Canonical connection.}
The canonical connection is related to the Levi-Civita connection by the 3-form $P$:
\begin{equation}
\begin{aligned}
 {}^P\Gamma_{\mu a}^b &= {}^{LC}\Gamma_{\mu a}^b + \frac{1}{m}P_{\mu ab} \\
 -{}^P\Gamma_{\mu a}^1 = {}^P\Gamma_{\mu 1}^a &= {}^{LC}\Gamma_{\mu 1}^a + P_{\mu 1a}.
\end{aligned}
\end{equation}
From the identities,
\begin{equation}
 \begin{aligned}
 P_{1ab}\gamma^{ab}\cdot\epsilon &= 2m\ii\gamma_1\cdot\epsilon \\
 P_{a1b}\gamma^{1b}\cdot\epsilon &= \ii\gamma_a\cdot\epsilon ,
 \end{aligned}
\end{equation}
and the Killing spinor equation, it follows that $\epsilon$ is parallel with respect to $\nabla^P$, and hence that
$\nabla^P$ has holonomy $\mbox{SU}(m)$.  Then $\nabla^P$ has to have a second parallel spinor $\tilde\epsilon$ as
discussed above.  From the identities,
\begin{equation}
 \begin{aligned}
 P_{1ab}\gamma^{ab}\cdot\tilde\epsilon &= (-1)^{m-1}2m\ii\gamma_1\cdot\epsilon \\
 P_{a1b}\gamma^{1b}\cdot\tilde\epsilon &= (-1)^{m-1}\ii\gamma_a\cdot\epsilon,
 \end{aligned}
\end{equation}
it follows that $\tilde\epsilon$ is a Killing spinor as well, with the same Killing constant as $\epsilon$ if and only
if $m$ is
odd.

The torsion of the connection $\nabla^P$ (which is metric-independent) can be calculated using the Cartan structure equation
\eqref{CSE}.  Thus,
\begin{equation}
\label{SE torsion}
\begin{aligned}
T^1 &= P_{1\mu\nu}e^\mu\wedge e^\nu \\
T^a &= \frac{m+1}{2m} P_{a\mu\nu}e^\mu\wedge e^\nu.
\end{aligned}
\end{equation}
The connection $\nabla^P$ is compatible with a whole family of metrics parametrised by a real constant $h$:
\begin{equation}
\label{SE metric}
 g_h = e^1e^1 + \exp(2h)\delta_{ab}e^ae^b.
\end{equation}
All of these metrics are Sasakian (up to homothety).  There are two special values of the parameter $h$.  The metric
with $h=0$ is special, because its Levi-Civita connection has a Killing spinor, it is Einstein, and its cone has
reduced holonomy.  On the other hand, the value
\begin{equation}
\label{SE special h}
 \exp(2h) = \frac{2m}{m+1}
\end{equation}
is special because this metric makes the torsion \eqref{SE torsion} of the canonical connection anti-symmetric (see also
\cite{FI01}).

Often a somewhat broader definition of the Sasaki-Einstein property is employed in the literature, which does not
guarantee the existence of Killing spinors on non-simply connected Sasaki-Einstein manifolds. On simply connected
manifolds the two definitions coincide \cite{BGbook, Sparks10}. 

 \paragraph{SU($m$)-instantons.}
The instanton condition 
\begin{equation}
 *F= -  *Q\wedge F =  - \frac { \eta \wedge \omega^{m-2}}{(m-2)!} \wedge F
\end{equation} 
is equivalent to $F\in \mathfrak{su}(m)$, which implies in particular $\eta \lrcorner F = \omega \lrcorner F=0$. 
Differentiating the instanton equation leads to the Yang-Mills equation
\begin{equation}
 \nabla^A \wedge * F + \frac {2\omega^{m-1}}{(m-2)!} \wedge F =0,
\end{equation} 
whose torsion term is proportional to $F\lrcorner (\eta \wedge \omega) $, and thus vanishes. Therefore the
instanton equation implies the Yang-Mills equation, confirming again proposition \ref{prop_YM+Spinors}.

\paragraph{Examples.}
In dimension 3 the only simply connected Sasaki-Einstein manifold is the sphere $S^3$, but already in dimension
5 a complete classification is missing. Many examples in arbitrary dimensions, including all homogeneous ones, can
be obtained from the following construction. Let $(N,g)$ be a $2m$-dimensional K\"ahler-Einstein manifold with positive
Ricci curvature Ric$^g=2mg$. Then there exists a principal U(1)-bundle on $N$ whose total space carries a
Sasaki-Einstein structure. Sasaki-Einstein manifolds obtained in this way are called regular; a generalization of this
construction to K\"ahler-Einstein orbifolds gives rise to quasi-regular Sasaki-Einstein manifolds. Homogeneous
Sasaki-Einstein manifolds are regular and can be obtained as circle bundles over generalized flag manifolds, including 
Hermitian symmetric spaces. Examples are
 \begin{itemize}
  \item odd-dimensional spheres $S^{2m+1} =$ SU$(m+1)/$SU($m$),
  \item Stiefel manifolds $V_2(\mathbb R^{m+1})=$ SO$(m+1)/$SO($m-1)$ (dimension $ 2m-1$),
  \item SO($2m)$/SU($m$) (dimension $m^2-m+1$), 
  \item Sp$(m)/$SU($m)$ (dimension $m^2+m+1$),
 \item $E_6/$SO(10) (dimension 33) and $E_7/E_6$ (dimension 55).
 \end{itemize}
 They are U(1)-bundles over irreducible compact Hermitian symmetric spaces, at least for $m$ large enough. Additional
homogeneous examples are obtained by allowing for a reducible base. Low-dimensional Sasaki-Einstein manifolds of this
type are the 7-dimensional spaces 
\begin{equation}
  Q(1,1,1) = {SU(2)^3 \over U(1)^2},
\end{equation} 
 with the U(1)$^2$-embedding orthogonal to the diagonal U(1)-subgroup, fibred over $\mathbb CP^1 \times \mathbb CP^1\times \mathbb CP^1$, and
 \begin{equation}
  M(3,2) = {SU(3)\times SU(2)\times U(1) \over SU(2)\times U(1) \times U(1)},
 \end{equation} 
 fibred over $\mathbb CP^2 \times \mathbb CP^1$ \cite{FFGRTZZ99}. The precise embedding of the subgroup for $M(3,2)$ is explained in \cite{Castellani}.


 Many non-regular and even irregular (non-quasi-regular) Sasaki-Einstein manifolds exist in dimension
$\geq 5$ \cite{BGbook, Sparks10}. For instance, $S^5$ and the Stiefel manifold
$S^2\times S^3$ carry several distinct quasi-regular non-regular Sasaki-Einstein structures. The same is true for the
connected sums $k(S^2\times S^3)$, where $k\geq 1$. Regular structures exist only up to $k=8$, and irregular structures
have been constructed on $S^2\times S^3$ \cite{GMSW04a}.

 In higher dimensions an interesting class of examples consists of exotic spheres. For
instance, all 28 smooth structures on $S^7$ admit several Sasaki-Einstein metrics \cite{BG05}. Families of Sasaki-Einstein manifolds in every odd dimension $\geq 5$ have been constructed in \cite{BG03, GMSW04b, CLPP05, LPV05}.

\subsection{3-Sasakian}\label{ssec:3Sgeom}

The subgroup $\mbox{Sp}(m)\subset\mbox{Spin}(4m+3)$ fixes $2m+2$ Dirac spinors.  The centraliser of $\mbox{Sp}(m)$ is
a subgroup $\mbox{Sp}(1)_1\times\mbox{Sp}(1)_2\subset\mbox{Spin}(4m+3)$, where Sp(1)$_1$, Sp($m)\subset $ Spin($4m$),
and Sp(1)$_2 =$ Spin(3). The $2m+2$ spinors transform in the
irreducible representations $\underline{m+1}$ of Sp(1)$_1$ and $\underline{2}$ of Sp(1)$_2$.  Of particular interest to
us will be the diagonal subgroup $\mbox{Sp}(1)_d$; the $2m+2$ spinors transform in the representation
\begin{equation}
 \underline{2}\otimes\underline{m+1} \cong \underline{m}\oplus\underline{m+2}
\end{equation}
of this subgroup.  An orthonormal basis for $\underline{m+2}$ will be labelled $\epsilon_A$, and for $\underline{m}$
$\tilde\epsilon_A$, where $A$ runs from 1 to $m$ or $m+2$ as appropriate.

A 3-Sasakian manifold is a $(4m+3)$-dimensional manifold with $m+2$ Killing spinors $\epsilon_A$.  Any such manifold
admits an Sp($m$)-structure.  There are $2m+2$ spinors $\epsilon_A,\tilde\epsilon_A$ which are parallel with respect to
any connection of holonomy Sp($m$); however, the additional spinors $\tilde\epsilon_A$ are not Killing spinors, as will
be proven below.

Any 3-Sasakian manifold admits a 2-sphere's worth of Sasaki-Einstein structures, which are rotated by the group $\mbox{Sp}(1)_d$.  The spinors that define these Sasaki-Einstein structures are highest weight vectors in the representation $\underline{m+2}$ of $\mbox{Sp}(1)_d$, and have stabiliser $\mbox{SU}(2m+1)$.  Associated to the Sasaki-Einstein structures are three 1-forms $\eta^\alpha$ and three 2-forms
$\omega^\alpha$.  In a local orthonormal frame $e^\alpha,e^a$, $\alpha=1,2,3,\ a=4,\dots,4m+3$, these can be written
\begin{equation}
 \begin{aligned}
  \eta^1&= e^1 & \omega^1 &= e^{45}+e^{67}+\dots +e^{4m\,4m+1}+e^{4m+2\,4m+3} \\
  \eta^2&= e^2 & \omega^2 &= e^{46}-e^{57}+\dots +e^{4m\,4m+2}-e^{4m+1\,4m+3} \\
  \eta^3&= e^3 & \omega^3 &= e^{47}+e^{56}+\dots +e^{4m\,4m+3}+e^{4m+1\,4m+2}.
 \end{aligned}
\end{equation}
The forms $\eta_\alpha,\omega_\alpha$ can be constructed as spinor bilinears in the highest weight Killing spinors as in
\eqref{SE bilinears}, and satisfy the differential identities,
\begin{equation}
 \begin{aligned}
  \dd\eta^\alpha &= \varepsilon_{\alpha\beta\gamma}\eta^\beta\wedge\eta^\gamma + 2\omega^\alpha \\
  \dd\omega^\alpha &= 2\varepsilon_{\alpha\beta\gamma}\eta^\beta\wedge\omega^\gamma.
 \end{aligned}
\end{equation}
The $\mbox{Sp}(1)_2$ rotates $\eta^\alpha$ and fixes $\omega^\alpha$, while Sp(1)$_1$
rotates
$\omega^\alpha$ and fixes $\eta^\alpha$, so that $\mbox{Sp}(1)_d$ rotates the Sasaki-Einstein structures.

The parallel forms $P',Q'$ satisfying \eqref{dPdQ} can be constructed as bilinears in the full set of $m+2$ Killing
spinors:
\begin{equation}
 \begin{aligned}
  P' &=& -\frac{\ii}{3!}\frac{1}{m+2} \sum_{A=1}^{m+2}  \big\langle \epsilon_A , \gamma_{\mu\nu\kappa} \epsilon_A
\big\rangle e^{\mu\nu\kappa} \\
  Q' &=& -\frac{1}{4!}\frac{1}{m+2} \sum_{A=1}^{m+2}  \big\langle \epsilon_A , \gamma_{\mu\nu\kappa\lambda} \epsilon_A
\big\rangle e^{\mu\nu\kappa\lambda}.
 \end{aligned}
\end{equation}
These do not coincide with the parallel forms $P,Q$ associated with the $\mbox{Sp}(m)$-structure.  The 3- and 4-forms
can be written in terms of the 1- and 2-forms as follows:
\begin{equation}
\begin{aligned}
 Q &= \frac{1}{6}\omega^\alpha\wedge\omega^\alpha \\
 Q' &= \frac{1}{6}\varepsilon_{\alpha\beta\gamma}\eta^{\alpha\beta}\wedge\omega^\gamma +
\frac{1}{6}\omega^\alpha\wedge\omega^\alpha \\
 P &= \frac13\eta^{123} + \frac13\eta^\alpha\wedge\omega^\alpha \\
 P' &= \eta^{123} + \frac13\eta^\alpha\wedge\omega^\alpha.
 \end{aligned}
\end{equation}
One can also construct 3- and 4-forms using the full set of $2m+2$ parallel spinors, and these are invariant under
$\mbox{Sp}(m)\times\mbox{Sp}(1)_1\times\mbox{Sp}(1)_2$:
\begin{equation}
\begin{aligned}
 Q &= -\frac{1}{4!}\frac{1}{2m+2}\left( \sum_{A=1}^{m+2}  \big\langle \epsilon_A , \gamma_{\mu\nu\kappa\lambda}
\epsilon_A \big\rangle + \sum_{A=1}^{m}  \big\langle \tilde\epsilon_A , \gamma_{\mu\nu\kappa\lambda} \tilde\epsilon_A
\big\rangle \right)e^{\mu\nu\kappa\lambda} \\
 \eta^{123} &= -\frac{\ii}{3!}\frac{1}{2m+2}\left( \sum_{A=1}^{m+2}  \big\langle \epsilon_A , \gamma_{\mu\nu\kappa}
\epsilon_A \big\rangle + \sum_{A=1}^{m}  \big\langle \tilde\epsilon_A , \gamma_{\mu\nu\kappa} \tilde\epsilon_A
\big\rangle \right)e^{\mu\nu\kappa}.
 \end{aligned}
\end{equation}

\paragraph{Canonical connection.}
The canonical connection is related to the Levi-Civita connection as follows:
\begin{equation}
\begin{aligned}
 -{}^{P}\Gamma_{\mu \alpha}^\nu = {}^{P}\Gamma_{\mu\nu}^\alpha &= {}^{LC}\Gamma_{\mu\nu}^\alpha + 3P_{\mu\nu\alpha} \\
 {}^{P}\Gamma_{\mu a}^b &= {}^{LC}\Gamma_{\mu a}^b
\end{aligned}
\end{equation}
From the identities,
\begin{equation}\label{3S:3-formspinorIdts}
\begin{aligned}
 P_{\alpha\beta\gamma}\gamma^{\beta\gamma}\cdot\epsilon_A &= \frac23\ii\gamma_\alpha\cdot\epsilon_A \\
 P_{ab\alpha}\gamma^{b\alpha}\cdot\epsilon_A &= \frac13\ii\gamma_a\cdot\epsilon_A,
\end{aligned}
\end{equation}
and the Killing spinor equation, it follows that the spinors $\epsilon_A$ are parallel with respect to $\nabla^P$, and
hence that $\nabla^P$ has holonomy $\mbox{Sp}(m)$.  Then $\nabla^P$ has to have in addition a set of parallel spinors
$\tilde\epsilon_A$ as discussed above.  They do not satisfy the identities \eqref{3S:3-formspinorIdts} however and hence
cannot be Killing spinors.

Up until now, the $m$ spinors $\tilde\epsilon_A$ have not
played a very prominent role in 3-Sasakian geometry, except in the case $m=1$ where upon a deformation of the
metric the single spinor $\tilde \epsilon$ can be made Killing. Since the other spinors $\epsilon_A$ are not
Killing for the deformed metric, the resulting space carries a strict nearly parallel $G_2$-structure \cite{Friedr97}.
See below for the deformation. With respect to the original 3-Sasakian metric, the structure defined by $\tilde
\epsilon$ in 7 dimensions is cocalibrated $G_2$ \cite{AF08}.

The torsion of $\nabla^P$ is calculated from \eqref{CSE}:
\begin{equation}
\label{3S torsion}
 \begin{aligned}
  T^{\alpha} &= 3P_{\alpha\mu\nu}e^{\mu\nu} \\
  T^a &= \frac32P_{a\mu\nu}e^{\mu\nu}.
 \end{aligned}
\end{equation}
The connection $\nabla^P$ is compatible with a whole family of metrics parametrised by a real constant $h$:
\begin{equation}
\label{3S metric}
 g_h = \delta_{\alpha\beta}e^\alpha e^\beta + \exp(2h)\delta_{ab}e^ae^b.
\end{equation}
Thus for 
\begin{equation}
\label{3S special h}
 \exp(2h)=2,
\end{equation}
the canonical connection has anti-symmetric torsion.  Two other special $h$-values are $h=0$ and $\exp(2h)=2m+3$; both
metrics are Einstein, but only the first is 3-Sasakian. In dimension 7 the metric with $\exp(2h)=5$ is nearly
parallel $G_2$ \cite{Friedr97}.

\paragraph{Sp($m$)-instantons.} Again there is no torsion in the Yang-Mills equation obeyed by Sp$(m$)-instantons. The
derivative of the instanton equation
 \begin{equation}
   \begin{aligned}{}
    *F &=- \frac 16 *(\omega^\alpha \wedge \omega^\alpha ) \wedge F
   \end{aligned}
 \end{equation} 
 gives
\begin{equation}
 \nabla^A\wedge *F\, \propto \, F\wedge *(\eta^\alpha\wedge \omega^\alpha).
\end{equation} 
Due to $\eta^\alpha\lrcorner F= \omega^\alpha \lrcorner F=0$ for $F\in \mathfrak{sp}(n)$ the right
hand side vanishes, confirming proposition \ref{prop_YM+Spinors}.

\paragraph{Examples.}
Homogeneous, simply connected 3-Sasakian manifolds are in a 1-1 correspondence with compact simple Lie groups:
\begin{equation}
 \begin{aligned}{}
  S^{4m+3}={Sp(m+1)\over Sp(m)}&, ~~~
{SU(m)\over S\big(U(m-2)\times U(1)\big)}, ~~~
{SO(m)\over SO(m-4)\times Sp(1)}, \\ 
 {G_2\over Sp(1)},\qquad &{F_4\over Sp(3)},\qquad {E_6\over
SU(6)},\qquad {E_7\over {\rm Spin}(12)},\qquad {E_8\over E_7}.
 \end{aligned}
\end{equation} 
Furthermore, there is only one family of non-simply connected homogeneous examples, given by the real projective
spaces $\mathbb{RP}^{4m+3} = S^{4m+3}/\mathbb Z_2$. Non-homogeneous 3-Sasakian manifolds can be constructed through a
reduction procedure \cite{BGbook, BG98}, and some examples are obtained as follows. Let $p\in \mathbb Z^{m+1}$ be such that
 \begin{equation}
   0 <p_1\leq \dots \leq p_{m+1}, \und \text{gcd}(p_i,p_j)=1\quad \forall i\neq j.
 \end{equation}  
 Define an action of U(1)$\times $U($m-1)$ on U$(m+1)$ through
 \begin{equation}
   (z,A) \cdot S = \text{diag}\big(z^{p_1} ,\dots, z^{p_{m+1}} \big)  \cdot S \cdot \left(\begin{array}{cc}
                                                                                      1_{2\times 2} & 0 \\
										      0 &  A  \\
                                                                                    \end{array}\right)
 \end{equation} 
 for $z\in S^1,\ A\in $ U($m-1)$ and $S\in $ U($m+1)$. Then the bi-quotient
 \begin{equation}
   S^m(p) =  \text U(m+1) \big/ \big(\text U(1)_p \times \text U(m-1)\big)
 \end{equation} 
  carries a 3-Sasakian structure. The dimension of $S^m(p)$ is $4m-1$, and for every $m$ the $S^m(p)$ give
infinitely many homotopy inequivalent simply connected compact inhomogeneous 3-Sasakian manifolds. In 7 dimensions the
$S^2(p)$ carry a second metric of positive sectional curvature. Equipped with this positive metric they are examples of
Eschenburg spaces \cite{Esch82}. Whether or not the Eschenburg metric coincides with the second nearly parallel $G_2$
metric that exists on every 3-Sasakian manifold is not known to us, but it seems at least plausible, given the fact that
the standard examples of nearly parallel $G_2$ manifolds all have positive sectional curvature.

 Similarly to the Sasaki-Einstein case, 3-Sasakian manifolds can be obtained as fibrations.
Let $(N,g)$ be a positive quaternionic K\"ahler manifold of dimension $4m$ and Ricci curvature Ric$^g=4(m+2)g$. Then
there exists a principal SO(3)-bundle over $N$ carrying a 3-Sasakian structure, which is regular by definition. A
generalization of this construction to quaternionic K\"ahler orbifolds gives rise to quasi-regular 3-Sasakian
manifolds, and it turns out that every 3-Sasakian manifold is quasi-regular.
Based on the LeBrun-Salamon conjecture that every positive quaternionic K\"ahler manifold is
symmetric \cite{LeBrunSalamon}, there is a conjecture that every regular 3-Sasaki manifold is homogeneous.

\subsection{Instantons}
 
The Riemann curvature form on a Riemannian manifold with reduced holonomy group $K\subset $ SO($n)$ has the following
properties
\begin{description}
 \item[] \ (1)\ $R$ takes values in the Lie algebra $\mathfrak{k}$, i.e.\ locally $R\in \mathfrak{k}\otimes\Lambda^2 \subset
\mathfrak{so}(n)\otimes\Lambda^2$.
 \item[] \ (2)\ $R$ has an interchange symmetry, i.e.\ $R_{\mu\nu\kappa\lambda}=R_{\kappa\lambda\mu\nu}$, where
$R_{\mu\nu\kappa\lambda}=g_{\mu\rho}R^\rho_{\nu\kappa\lambda}$ and
$R^\mu_\nu=\sfrac{1}{2}R^\mu_{\nu\kappa\lambda}e^\kappa\wedge e^\lambda$.
\end{description}
Together these imply that locally $R\in\mathfrak{k}\otimes\mathfrak{k}$, so that $R$ solves the instanton equation
\eqref{instcond2}.  On a Riemannian manifold with a $K$-structure an arbitrary connection with holonomy group $K$ has
the first property, but due to the existence of torsion the second property may fail.  The following proposition shows
that the canonical connection has both properties:
\begin{prop}
\label{prop:TorsionExSymmetry}
 Let $\nabla^t$ be a metric-compatible connection with totally anti-symmetric torsion $T^\mu=te^\mu\lrcorner P$ for some
3-form $P$ and real parameter $t$.  Suppose that when $t=1$, $P$ is parallel, that is, $\nabla^1P = 0$.  Then the
curvature of $\nabla^t$ satisfies property (2) for all $t$.
\end{prop}
The most important case $t=1$ of this proposition appeared earlier in \cite{Agricola06}; we give a proof here for the
sake of completeness.
\begin{proof}[Proof of Prop. \ref{prop:TorsionExSymmetry}]
 Let $e^\mu$ be a local orthonormal frame for the cotangent bundle, and let ${}^t\Gamma^\mu_\nu$ be the matrix of
1-forms which defines the connection $\nabla^t$.  Applying the exterior derivative to the Cartan structure equation \eqref{CSE} for $\nabla^t$ yields the 1st Bianchi identity:
 \begin{equation}
 \label{BI1}
  0 = \dd^2 e^\mu = -{}^tR^\mu_\nu\wedge e^\nu + \dd {}^tT^\mu + {}^t\Gamma^\mu_\nu\wedge {}^tT^\nu,
 \end{equation}
 where ${}^tR^\mu_\nu = \dd{}^t\Gamma^\mu_\nu + {}^t\Gamma^\lambda_\nu\wedge{}^t\Gamma^\mu_\lambda$ is the curvature. 
Thus in order to understand the conditions imposed on the curvature by the Bianchi identity, we need to first evaluate
$\dd {}^tT^\mu + {}^t\Gamma^\mu_\nu\wedge {}^tT^\nu$.
 
 We consider first the special case $t=1$.  The 3-form $P=\sfrac{1}{6}P_{\mu\nu\lambda}e^{\mu\nu\lambda}$ is parallel,
and in components this means that
 \begin{equation}
  0 = \dd P_{\mu\nu\lambda} - {}^1 \Gamma^\rho_\mu P_{\rho\nu\lambda} - {}^1 \Gamma^\rho_\nu P_{\mu\rho\lambda} - {}^1
\Gamma^\rho_\lambda P_{\mu\nu\rho} .
 \end{equation}
 Now ${}^1  T^\mu=\sfrac{1}{2}P_{\mu\nu\lambda}e^{\nu\lambda}$, and it follows that
 \begin{equation}
  \dd{}^1  T^\mu + {}^1 \Gamma^\mu_\nu\wedge{}^1  T^\nu = -\frac{1}{2}
P_{\mu\nu\rho}P_{\rho\kappa\lambda}e^{\nu\kappa\lambda}.
 \end{equation}
 The Christoffel symbols in the general case are related to those in the case $t=1$ by
 \begin{equation}
  {}^t\Gamma^\mu_\nu = {}^1 \Gamma^\mu_\nu + \frac{1-t}{2} P_{\mu\nu\lambda}e^\lambda,
 \end{equation}
 as follows from the Cartan structure equation (\ref{CSE}).  Therefore in general we have
 \begin{equation}
  \dd {}^tT^\mu + {}^t\Gamma^\mu_\nu\wedge {}^tT^\nu = -\frac{t(t-3)}{4}
P_{\mu\nu\rho}P_{\rho\kappa\lambda}e^{\nu\kappa\lambda}.
 \end{equation}
 
 Now we are ready to understand the implications of the first Bianchi identity.  We define a tensor
$C_{\mu\nu\kappa\lambda}$ by
 \begin{equation}
 \label{C tensor}
  C_{\mu\nu\kappa\lambda} = {}^tR_{\mu\nu\kappa\lambda} + \frac{t(t-3)}{4} P_{\mu\nu\rho}P_{\rho\kappa\lambda}.
 \end{equation}
 This tensor satisfies the following identities (the last of which follows from \eqref{BI1}):
 \begin{eqnarray}
  C_{\mu\nu\kappa\lambda} + C_{\nu\mu\kappa\lambda} &=& 0 \\
  C_{\mu\nu\kappa\lambda} + C_{\mu\nu\lambda\kappa} &=& 0 \\
  C_{\mu\nu\kappa\lambda} + C_{\mu\kappa\lambda\nu} + C_{\mu\lambda\nu\kappa} &=& 0.
 \end{eqnarray}
 It follows that $C_{\mu\nu\kappa\lambda}$ has the interchange symmetry:
 \begin{equation}
  C_{\mu\nu\kappa\lambda} = C_{\kappa\lambda\mu\nu}.
 \end{equation}
 It then follows straightforwardly from (\ref{C tensor}) that ${}^tR_{\mu\nu\kappa\lambda}$ also has the interchange
symmetry.
\end{proof}

\begin{coroll}
\label{cor:caninst}
 Let $M$ be a manifold with real Killing spinor, then its canonical connection $\nabla^P$ is an instanton.  In the
special cases of Sasaki-Einstein manifolds and 3-Sasakian manifolds, the instanton equation is solved for all values of
the metric parameter $h$.
\end{coroll}
\begin{proof}
 That $\nabla^P$ solves the instanton equation for the special values \eqref{SE special h}, \eqref{3S special h} of $h$
is immediate.  The proof is completed by the observation that the instanton equation $F\in\mathfrak{k}$ is
$h$-independent.
\end{proof}

Apart from manifolds with real Killing spinor, the other main examples of manifolds with a canonical connection are reductive homogeneous manifolds (or coset spaces).  On such manifolds our notion of canonical connection coincides with the usual notion of canonical connection \cite{Kobayashi-Nomizu1}.  In particular, the canonical connection on any coset space is an instanton, as was previously shown in \cite{HP10}.

We end this section with some comments on the relation between our canonical connection and the ``characteristic
connection'' introduced in \cite{FI01,Agricola06}.  A characteristic connection on a manifold with $K$-structure is a
connection with holonomy $K$ and totally anti-symmetric torsion, if such a connection exists.  Nearly K\"ahler manifolds
and nearly parallel $G_2$-manifolds admit a unique characteristic connection, and this coincides with our canonical
connection.  

On Sasaki-Einstein manifolds there is a unique characteristic connection, and it has holonomy $\mbox{U}(m)$ and is totally
anti-symmetric torsion with respect to the Einstein metric.  On the other hand, the canonical connection is
characterised by holonomy group $\mbox{SU}(m)$, and torsion which is totally anti-symmetric with respect to one of the
metrics compatible with the $\mbox{SU}(m)$-structure.  Thus the canonical connection differs from the characteristic
connection by satisfying a stronger holonomy condition, and a weaker torsion condition.

For the purposes of the present article, the canonical connection has two main advantages over the characteristic
connection.  Firstly, the canonical connection satisfies the instanton equation, while on a Sasaki-Einstein manifold the
characteristic connection does not.  And secondly, a canonical connection exists on a 3-Sasakian manifold, whereas it can be
proven that no characteristic connection exists in this case.

On nearly K\"ahler and nearly parallel $G_2$-manifolds the canonical connection is the same as a characteristic
connection, so is unique \cite{FI01}.  On Sasaki-Einstein manifolds the canonical connection coincides with the
characteristic connection for the special $h$-value \eqref{SE special h}.  Now for each value of $h$ there exists a
unique characteristic connection with holonomy U($m$) \cite{FI01}, and this has holonomy SU($m$) only when $h$ satisfies
\eqref{SE special h}.  Therefore the canonical connection of a Sasaki-Einstein manifold is also unique.  We have not
investigated whether 3-Sasakian manifolds also have a unique canonical connection, but clearly this is an interesting
question for further investigation.

\section{Instantons on the cone}
\label{sec:cone}

Having constructed examples of instantons on manifolds $M$ with real Killing spinor, we now turn our attention to their
cones.  It will actually prove more convenient to study the instanton equation on the cylinder $Z=\RR\times M$, equipped
with metric
\begin{equation}
\label{cylinder metric}
 g_Z = \dd\tau^2 + g_h.
\end{equation}
In the Sasaki-Einstein and 3-Sasakian cases $g_h$ is taken to be the $h$-dependent metric \eqref{SE metric}, \eqref{3S
metric} (with $h$ promoted to a function of $\tau$), and in the nearly K\"ahler and nearly parallel $G_2$ cases
$g_h=\delta_{ab}e^ae^b$ is the usual Einstein metric.  The cylinder inherits a $K$-structure from $M$, and this can be
lifted to a $G$-structure, where
$G=\mbox{Spin}(7)$, $G_2$, $\mbox{SU}(m+1)$ or $\mbox{Sp}(m+1)$ when $M$ is nearly parallel $G_2$, nearly K\"ahler,
Sasaki-Einstein or 3-Sasakian.  The instanton equation on the cylinder is $F\in\mathfrak{g}$, or equivalently
\begin{equation}
\label{cone inst eq}
 \ast F = - \ast Q_Z\wedge F,
\end{equation}
where $Q_Z$ is the Casimir 4-form associated to the $G$-structure on the cylinder.  Since the instanton equations are
conformally invariant, and the cylinder metric is conformal to the the cone metric, instantons on the cylinder will also
be instantons on the cone.

There are two obvious examples of instantons on the cylinder (or cone): the Levi-Civita connection $\nabla^C$ on the
cone is an instanton, because the cone is a manifold of special holonomy, and the canonical connection $\nabla^P$ on $M$
lifts to an instanton on the cylinder.  Both of these connections have holonomy contained in the structure group $G$ of
the cylinder.  The instantons constructed in this section also have holonomy group $G$.  They interpolate between the
Levi-Civita and canonical connections: at the apex $\tau=-\infty$ they agree with the Levi-Civita connection, and at the
boundary $\tau=\infty$ they agree with the canonical connection.  The instantons depend on a single parameter $\tau_0$:
this is a translational parameter from the point of view of the cylinder, or a scale parameter from the point of view of
the cone.

If $M$ is a sphere its cone can be completed by adding a point at the apex $\tau=-\infty$, forming the manifold $\RR^{n+1}$.  The instantons that we construct asymptote to the Levi-Civita connection on $\RR^{n+1}$ as $\tau\to-\infty$, and it follows that they can be extended over the apex.  Thus we obtain instantons on Euclidean spaces.  The zero-size limits are interesting, because these give examples of singularity formation.  In fact, the $\tau_0\to-\infty$ limits of our instantons are examples of ``tangent connections'' in the language of Tian \cite{Tian00}.

\subsection{Nearly K\"ahler and nearly parallel $G_2$}

Nearly parallel $G_2$-manifolds and nearly K\"ahler 6-manifolds are sufficiently similar to be treated in a unified way.
 In both cases the Casimir 4-form on the cylinder is
\begin{equation}
\label{NKNP Q}
 Q_Z = \dd\tau\wedge P + Q.
\end{equation}
The canonical connection lifts to a connection on $TZ$ with holonomy group $K=G_2$ or SU(3).  Our ansatz for a
connection on the cylinder $\RR\times M$ will be a perturbation of the canonical connection.  This perturbation will be
made using a parallel section of $T^\ast Z\otimes \mbox{End}(TZ)$, in such a way that the gauge group of the perturbed
connection will be $G=\mbox{Spin}(7)$ or $G_2$.

Recall that $\mathfrak{g}=\mathfrak{k}\oplus\mathfrak{m}$, and that $K$ acts irreducibly on $\mathfrak{k}$ and its
$n$-dimensional orthogonal complement $\mathfrak{m}$.  The bundle over $M$ defined by this $K$-representation
pulls back
to the subbundle of $\mbox{End}(TZ)$ with fibre $\mathfrak{g}$.  Similarly, the representation of $K$ that defines
$T^\ast Z$ splits into two irreducible pieces of dimensions $n$ and 1, and the $n$-dimensional piece can be identified
with $\mathfrak{m}^\ast$.  It follows that the tensor product of these two representations has a 1-dimensional trivial
subrepresentation, and hence that the associated bundle over $M$ admits a parallel section, which we pull back to
$T^\ast Z\otimes \mbox{End}(TZ)$. 

To make this parallel section more explicit, we now choose a local frame $e^a$, for $T^\ast M$ so that the 3-form $P$
takes its standard form, as described in the preceding section.  We extend this to a local frame for $T^\ast Z$ by
defining $e^0=\dd\tau$.  Then there is an associated basis $I_a$, $a=1,\ldots,n$ for
$\mathfrak{m}\subset\mathfrak{g}\subset\mathfrak{so}(n+1)$.  Since these are $(n+1)$-dimensional matrices we can attach
matrix indices $\mu,\nu=0\ldots n$, so that the generators are $I_{a\mu}^\nu$.  These matrices can be written explicitly
as follows:
\begin{equation}
 -I_{ab}^0=I_{a0}^b=\delta_a^b,\quad I_{ab}^c=-\frac{1}{\rho}P_{abc},
\end{equation}
where $\rho=2,3$ in the cases $n=6,7$.  One way to see that these matrices belong to
$\mathfrak{g}\subset\mathfrak{so}(n+1)$ is to note that the 2-forms,
\begin{equation}
\label{SD 2-form}
 e^{0a} - \frac{1}{2\rho}P_{abc}e^{bc},
\end{equation}
solve the instanton equation \eqref{cone inst eq} on the cylinder, so belong to $\mathfrak{g}\subset\Lambda^2$.  The
generators $I_a$ are the images of these 2-forms under the metric-induced isomorphism
$\Lambda^2\cong\mathfrak{so}(n+1)$.  The parallel section that we will use in our ansatz is simply $e^aI_a$.

The matrices $I_a$ are orthonormal with respect to a multiple of the Cartan-Killing form on $\mathfrak{g}$, and we
extend them to a basis for $\mathfrak{g}$ using an orthonormal basis $I_i$ for $\mathfrak{k}$.  Clearly
$I_{ia}^0=-I_{i0}^a=0$.  The structure constants satisfy
\begin{equation}
\label{NKNP structure constants}
 f_{ib}^a = I_{ib}^a ,\quad f^a_{bc} = -\frac{2}{\rho}P_{abc}.
\end{equation}
Here the first equality merely expresses the fact that $\mathfrak{m}$ and $\RR^n$ are isomorphic as $\mathfrak{k}$
representations.

The ansatz for a connection on the cylinder may now be written
\begin{equation}
\label{NKNP inst ansatz}
 \nabla^A = \nabla^P + \psi(\tau)e^aI_a.
\end{equation}
When $\psi(\tau)=1$, $\nabla^A$ is in fact the Levi-Civita connection $\nabla^C$ on the cone, so that the ansatz could
be rewritten
\begin{equation}
\label{NKNP inst ansatz 3}
 \nabla^A = \nabla^P + \psi(\tau)(\nabla^C-\nabla^P).
\end{equation}
To prove this one needs to show that the connection with $\psi(\tau)=1$ is torsion-free when acting on an orthonormal
frame for the cone metric.  This will be done in the next section.  In the nearly K\"ahler case there is also a parallel
section $e^a\omega_{ab}I_b$, however as in \cite{HILP10}, the additional instantons obtained by including this in the
ansatz \eqref{NKNP inst ansatz} are related to the ones for our simpler ansatz by a rotation by $\pm 2\pi
/3$ in the parameter plane, so we omit this additional term.

Now we will calculate the curvature of $\nabla^A$.  Note that
\begin{eqnarray}
 \dd (e^aI_a) + {}^{P}\Gamma\wedge I_ae^a+ I_ae^a\wedge{}^{P}\Gamma &=& I_a \dd e^a + {}^{P}\Gamma_a^bI_b \wedge e^a \\
 &=& I_aT^a \\
 &=& \frac{1}{\rho} P_{abc}I_a e^{bc}.
\end{eqnarray}
Here the first equality follows from \eqref{NKNP structure constants}: we may write ${}^P\Gamma={}^P\Gamma^iI_i$, so that ${}^P\Gamma^i[I_i,I_a]={}^P\Gamma^iI_{ia}^bI_b={}^P\Gamma_a^bI_b$.  The second equality follows from the Cartan structure equation \eqref{CSE}, and in the third equality we have inserted the torsion \eqref{NP torsion}, \eqref{NK torsion}.  So the curvature of the connection is
\begin{eqnarray}
 F = R^P + \frac{1}{2}\psi^2f_{ab}^ie^{ab}I_i + \dot \psi e^{0a}I_a + \frac{1}{\rho}(\psi-\psi^2) P_{abc}e^{bc}I_a,
\end{eqnarray}
where $R^P$ is the curvature of $\nabla^P$.

Now we consider whether $F$ solves the instanton equation.  We already know that $R^P$ does.  It is also not hard to see
that the term $f_{ab}^ie^{ab}$ also solves the instanton equation.  The map $I_i\mapsto f_{ia}^b$ describes the
embedding $\mathfrak{k}\mapsto\mathfrak{so}(n)$, so for each $i$, the 2-form $f_{ab}^ie^{ab}$ lies in the subspace
$\mathfrak{k}\subset\Lambda^2$.  Alternatively, one only needs to note that $R^P + \frac{1}{2}f_{ab}^ie^{ab}I_i$
is the curvature Levi-Civita connection on the cone, and hence an instanton.

Thus $F$ is an instanton if and only if the $I_a$ terms solve the instanton equation, and from equation (\ref{SD
2-form}) one easily sees that this happens exactly when
\begin{equation}
\label{NKNP inst}
 \dot \psi = 2(\psi^2-\psi).
\end{equation}
The solution of this differential equation is
\begin{equation}\label{NKNP instSol}
 \psi = \left( 1 + e^{2(\tau-\tau_0)} \right)^{-1}.
\end{equation}
The limit $\tau_0\to-\infty$ is the original connection $\nabla^P$ on the cylinder, and the limit $\tau_0\to\infty$ is
the Levi-Civita connection on the cone.  When $M=S^6$ or $S^7$ this construction reproduces the instantons of
\cite{FN84,FN85,IP92,GN95} on $\RR^7,\RR^8$.  Also, when $M=S^6$, this construction is equivalent to one given in \cite{HILP10}; however,
for the other nearly K\"ahler and coset spaces this construction differs from \cite{HILP10} (see also \cite{Gemmer11}).  A more general class of instantons have been constructed on the cones over Aloff-Wallach manifolds in \cite{HILP11}.

\subsection{Sasaki-Einstein}

The Casimir 4-form on the cylinder over a Sasaki-Einstein manifold depends on the metric parameter $h$, which is
promoted to a function of $\tau$:
\begin{equation}
\label{SE Q}
 Q_Z = e^{2h(\tau)}\dd\tau\wedge P + e^{4h(\tau)}Q.
\end{equation}
We construct instantons on Sasaki-Einstein manifolds by the same method as in the nearly K\"ahler and nearly parallel
$G_2$-cases.  The main deviation is that now there are 3 parallel sections rather than 1.

Again, we write $\mathfrak{g}=\mathfrak{k}\oplus\mathfrak{m}$, where $\mathfrak{g}=\mathfrak{su}(m+1)$,
$\mathfrak{k}=\mathfrak{su}(m)$, and $\mathfrak{m}$ is the $(2m+1)$-dimensional space orthogonal to $\mathfrak{k}$.  Once
again, $\mathfrak{m}^\ast$ is isomorphic to the $(2m+1)$-dimensional orthogonal representation of $\mathfrak{su}(m)$ that
defines the cotangent bundle $T^\ast M$.  The tensor product of these two representations contains a 3-dimensional
trival sub-representation, which gives 3 parallel sections which we pull back to $T^\ast Z\otimes \mbox{End}(TZ)$.

To construct the parallel sections, we assume that a local orthonormal basis $e^1,e^a$ for $T^\ast M$ has been chosen so
that the parallel forms take their standard forms, and also set $e^0=\dd\tau$.  The generators of $\mathfrak{k}$ will be
denoted $I_i$ and the additional generators of $\mathfrak{m}$ associated to the frame $e^1,e^a$ will be denoted
$I_1,I_a$.  Written as matrices, these have the following non-vanishing elements
\begin{equation}
\begin{aligned}
 I_{ia}^b =& f_{ia}^b,  \\
 I_{1a}^b =& -\sfrac{1}{m}\omega_{ab}, & -I_{11}^0=&I_{10}^1 = 1, \\
 -I_{ab}^0=& I_{a0}^b = \delta_a^b, & -I_{ab}^1=&I_{a1}^b = \omega_{ab}.
\end{aligned}
\end{equation}
In particular, the matrices $I_1,I_a$ are the images in $\mathfrak{so}(2m+2)$ of the anti-self-dual 2-forms,
\begin{equation}
e^{01}-\frac{\exp(2h)}{m}\omega,\quad \exp(h)(e^{0a}+\omega_{ab}e^{1b}).
\end{equation}
With this choice of basis, the structure constants satisfy
\begin{equation}
f_{ab}^1=-2P_{ab1},\quad f_{ab}^c=0,\quad f_{1a}^b=-\frac{m+1}{m}P_{1ab}.
\end{equation}
Notice the similarity with the formulae \eqref{SE torsion} for the torsion.

The three parallel sections are $e^aI_a$, $e^a\omega_{ab}I_b$ and $e^1I_1$, so the natural ansatz for a connection is
\begin{equation}
\label{SE inst ansatz}
 \nabla^A = \nabla^P + \chi(\tau)e^1I_1 + \psi(\tau)e^aI_a + \tilde\psi(\tau)e^a\omega_{ab}I_b,
\end{equation}
with $\chi,\psi,\tilde\psi$ real functions of $\tau$.  It can be shown that the instanton equation implies that the
argument of $\psi+\ii\tilde\psi$ is constant and that the instanton equation is invariant under phase rotations of this
complex variable.  Therefore one can always fix $\tilde\psi=0$, and we will do so here in order to simplify the
presentation.

The curvature of this connection is
\begin{eqnarray}
 F &=& R^P + \frac{1}{2}\psi^2f_{ab}^ie^{ab}I_i + \dot\chi e^{01}I_1 + \dot\psi e^{0a}I_a\\
 \nonumber && + (\chi-\psi^2) P_{ab1}e^{ab}I_1 + \frac{m+1}{m}(\psi-\psi\chi)P_{1ba}e^{1b}I_a.
\end{eqnarray}
Once again, $R^P$ is the curvature of $\nabla^P$ so solves the instanton equation.  The term $\psi^2f_{ab}^ie^{ab}I_i$
also solves the instanton equation, as can be shown either by a direct argument, or by using the fact (to be proved in
the
next section) that the connection with $\psi=\chi=1$ is the Levi-Civita connection $\nabla^C$ on the cone.  Therefore the instanton
equation is equivalent to
\begin{eqnarray}
 \label{SE gaugino 1}
 \dot\chi &=& 2me^{-2h}(\psi^2-\chi) \\
 \label{SE gaugino 2}
 \dot\psi &=& \frac{m+1}{m}\psi(\chi-1).
\end{eqnarray}
The ansatz \eqref{SE inst ansatz} and the associated instanton equations \eqref{SE gaugino 1}, \eqref{SE gaugino 2} are
equivalent to those given in \cite{Correia09}.

\begin{figure}[ht]
 \psfrag{p}{$\psi$}
 \psfrag{c}{$\chi$}
 \psfrag{mequ1}{$m=1$}
 \psfrag{mequ2}{$m=2$}
 \psfrag{mequ4}{$m=4$}
 \psfrag{mequ8}{$m=8$}
 \psfrag{mequinf}{$m=\infty$}
 \epsfig{ file = 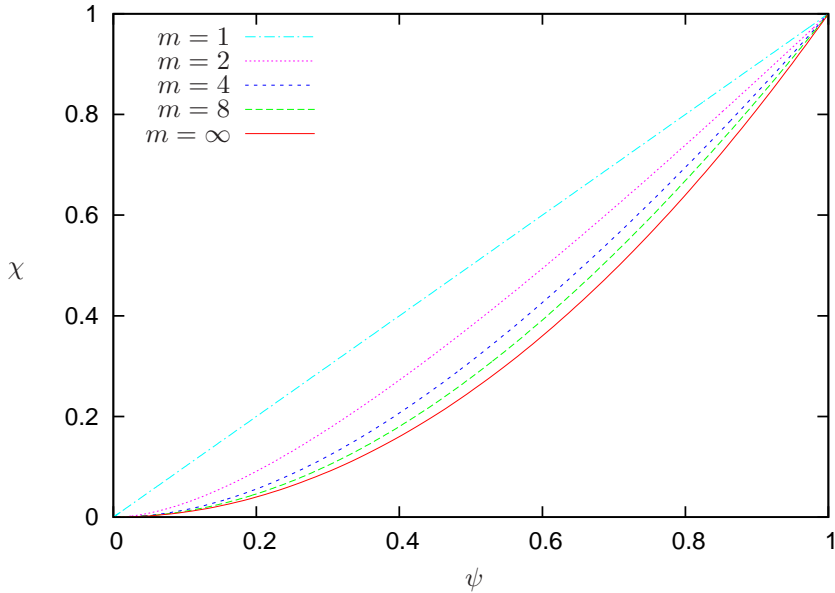, scale = 1.3 }
 \caption{Instantons on the cone over Sasaki-Einstein manifolds, plotted in the $\psi,\chi$ plane.  The dashed and dotted
curves are instantons with $h=0$ and $m=1,2,4,8$.  For $m=1$ we have $\psi =\chi$, and as $m$ increases the solutions get
closer to the limiting solid curve $\psi^2=\chi$.}
 \label{fig1}
\end{figure}

The flow equations \eqref{SE gaugino 1}, \eqref{SE gaugino 2} have two fixed points at $(\psi,\chi)=(0,0)$ and $(1,1)$
corresponding to the instantons $\nabla^P$ and $\nabla^C$.  The first critical point is stable and the second
semi-stable, so assuming that solutions to these equations exist for all time, there is a 1-parameter family of solutions interpolating from the second to the first (at least for reasonable choices of $h$).  If $h$ is independent of $\tau$ the parameter may be interpreted as a translational parameter $\tau_0$.  When $h=0$ and $m=1$ there is an exact solution,
\begin{equation}
\label{SE instSol}
 \psi = \chi = \left(1+e^{2(\tau-\tau_0)}\right)^{-1},
\end{equation}
which is just the BPST instanton on $\RR^4$ \cite{BPST75}.  For $m>1$ there are similar exact solutions with $\chi=\psi$ when $e^{2h}=2m^2/(m+1)$.  Exact solutions of this type were previously constructed on homogeneous spaces (including homogeneous Sasaki-Einstein manifolds) in \cite{ILPR09}.

However, the most interesting choice for $h$ is $h=0$, corresponding to
the Einstein metric.  For $m>1$ solutions can be found only numerically, and a sample are depicted in figure \ref{fig1}.  These solutions were constructed using a Runge-Kutta algorithm, and the boundary condition $(\psi,\chi)\to(1,1)$ as $\tau\to-\infty$ was imposed by shooting from the line $\psi+\chi=1$.
 There is however an exact solution in the $m\to\infty$ limit: in this limit, equation \eqref{SE gaugino 1} simplifies
to $\psi^2=\chi$ and equation \eqref{SE gaugino 2} becomes
\begin{equation}
 \dot\psi = 2\psi(\psi^2-1).
\end{equation}
This is solved by $\psi=\left( 1 + e^{2(\tau-\tau_0)}) \right)^{-1/2}$.

Of particular interest are the cases where $M=S^{2m+1}$.  In these cases the cone metric extends smoothly over the apex
$\tau=-\infty$ to form the manifold $\RR^{2m+2}$.  The instantons also extend over the apex, since at $\tau=-\infty$
they coincide with the Levi-Civita connection on $\RR^{2m+2}$, which does extend over the apex.  Thus the instantons
that we have constructed include a new family of instantons on even-dimensional Euclidean spaces.  The $\tau_0\to-\infty$ limit of the instanton on $\RR^6$ is the example of a ``tangent connection'' given in \cite{Tian00}.

\subsection{3-Sasakian}

The Casimir 4-form on the cylinder is
\begin{equation}
\label{3S Q}
  Q_Z  = \frac{1}{6} \Big(e^{4h}\omega^\alpha \wedge \omega^\alpha +e^{2h}\varepsilon_{\alpha\beta\gamma } \omega^\alpha
\wedge \eta^{\beta\gamma} + 2e^{2h}\dd\tau \wedge \eta^\alpha \wedge \omega^\alpha 
       +6 \dd\tau \wedge \eta^{123} \Big),
\end{equation} 
where once again we allow $h$ to depend on $\tau$.  As above, we write $\mathfrak{g}=\mathfrak{k}\oplus\mathfrak{m}$,
with $\mathfrak{g}=\mathfrak{sp}(m+1)$, $\mathfrak{k}=\mathfrak{sp}(m)$, and $\mathfrak{m}$ is the $(4m+3)$-dimensional
space orthogonal to $\mathfrak{k}$.  Once again, $\mathfrak{m}^\ast$ is isomorphic to the $(4m+3)$-dimensional orthogonal
representation of $\mathfrak{sp}(m)$ that defines $T^\ast M$.

We assume that a local orthonormal basis $e^\alpha,e^a$ for $T^\ast M$ has been chosen so that the parallel forms take
their standard forms, and also set $e^0=\dd\tau$.  The generators of $\mathfrak{k}$ will be denoted $I_i$ and the
additional generators of $\mathfrak{m}$ associated to the frame $e^\alpha,e^a$ will be denoted $I_\alpha,I_a$.  The
non-vanishing components of these matrices are
\begin{equation}
\begin{aligned}
 I_{ia}^b =& f_{ia}^b, \\
 I_{\alpha0}^\beta =& \delta_{\alpha}^\beta, & I_{\alpha \beta}^\gamma =& -\varepsilon_{\alpha\beta\gamma},  \\
 I_{a0}^b =& \delta_a^b, & I_{ab}^\alpha =& -\omega_{ab}^\alpha. \\
\end{aligned}
\end{equation}
In particular, the matrices $I_\alpha,I_a$ are the images in $\mathfrak{so}(4m+4)$ of the anti-self-dual 2-forms,
\begin{equation}
 e^{0\alpha} -\frac 12 \varepsilon_{\alpha\beta\gamma} e^{\beta\gamma},\quad \exp(h)( e^{0a} + \omega^\alpha_{ab}
e^{\alpha b}).
\end{equation}
The Lie algebra structure constants satisfy
\begin{equation}
 \begin{aligned}{}
  f_{\beta\gamma}^\alpha &= -2 \varepsilon_{\alpha\beta\gamma}, \quad
  f_{ab}^\alpha  =  -2\omega^\alpha_{ab}, \quad
  f_{\alpha a}^b = -\omega^\alpha_{ab} ,
 \end{aligned}
\end{equation} 
which should be compared to the torsion \eqref{3S torsion}.

There are 2 matrix-valued forms which are parallel with respect to connections with holonomy
$\mbox{Sp}(1)_d\times\mbox{Sp}(m)$.  We use both to make an ansatz for a connection:
\begin{equation}
\label{3S inst ansatz}
 \nabla^A = \nabla^P + \chi(\tau ) e^\alpha I_\alpha + \psi (\tau) e^a I_a,
\end{equation}
with $\chi,\psi$ real functions of $\tau$.  Using the above result \eqref{3S torsion} for the
canonical torsion, we obtain for the curvature of the connection:
\begin{equation}
\begin{aligned}{}
    F&= R^P + \frac 12 \psi ^2 f^k_{ab} e^{ab} I_k \\
   &\quad + \Big( \dot \chi e^{0\alpha} +2(\chi -\psi^2) \omega^\alpha +\chi(1-\chi) \varepsilon_{\alpha\beta\gamma}
e^{\beta\gamma} \Big) I_\alpha \\
  &\quad +\Big(\dot \psi e^{0a} -\psi (1-\chi) \omega^\alpha_{ab} e^{\alpha b}\Big) I_a.
\end{aligned}
\end{equation} 
Once more, the connection with $\chi=\psi=1$ is the Levi-Civita connection on the cone.  The first two terms solve the
instanton equation, using the fact that the canonical connection and the Levi-Civita connection on the cone are
instantons.  Thus the instanton equation reduces to
\begin{eqnarray}
\label{3SInstNecCond1}
 0 &=& \chi - \psi^2, \\
\label{3SInstNecCond2}
 \dot \chi &=& 2 \chi ( \chi -1) ,\\
\label{3SInstNecCond3}
 \dot \psi &=& \psi (\chi -1),
\end{eqnarray}
which are independent of $h(\tau)$.  Note that these are 3 equations for 2 unknown functions, so naively one would not
expect to find any solutions.  However, in the case at hand the condition (\ref{3SInstNecCond1}) is conserved by the
flow described by the other two equations, so solutions can be found.  They are:
\begin{eqnarray}
\label{3SInstSoln1}
 \chi &=& \Big( 1 + e^{2(\tau-\tau_0)} \Big)^{-1} ,\\
\label{3SInstSoln2}
 \psi &=& \pm \Big(1 + e^{2(\tau-\tau_0)} \Big) ^{-1/2}.
\end{eqnarray}
When $M=S^{4m+3}$ our construction produces an instanton on $\RR^{4m+4}$ (which is the BPST instanton when $m=0$). 
When $m\geq1$ these instantons probably coincide with the quaternionic instantons constructed in \cite{CGK85,BIL08},
however a direct comparison is not possible since curvatures were not calculated in \cite{CGK85,BIL08}.

\subsection{Gradient flows}

The instanton equations on the cylinder have an interesting interpretation as gradient flow equations.  Suppose that $A$
is a gauge field on the cylinder, and that a gauge has been chosen in which $A_\tau=0$, so that $A$ can be
thought of as a $\tau$-dependent gauge field on $M$, with curvature $\dd\tau\wedge\dot A + F$.  The instanton equation
\eqref{cone inst eq} is equivalent to
\begin{eqnarray}
\label{grad flow}
 \ast\dot A &=& - \ast P' \wedge F \\
\label{grad constraint}
 \ast F &=& -\dot A \wedge \ast P' - \ast Q'\wedge F,
\end{eqnarray}
where all Hodge stars are taken with respect to the metric on $M$.  Now consider the Chern-Simons functional,
\begin{equation}
 W = \int_M \Tr\left(A\wedge\dd A + \frac23 A\wedge A\wedge A\right)\wedge\ast P'=\frac{1}{n-3}\int_M\Tr\left( F\wedge
F\right)\wedge\ast Q'.
\end{equation}
This functional is gauge-invariant when $n>3$ (on $S^3$ it is gauge invariant modulo $\mathbb{Z}$).  The space of all
connections $A$ on $M$ can be given an $L^2$ metric, and the gradient flow equation for $W$ is then the first equation
\eqref{grad flow}.

If $M$ is a nearly parallel $G_2$-manifold, the following identity holds for any 2-form $F$:
\begin{equation}
 Q\wedge\ast(Q\wedge F) = \ast Q\wedge F + \ast F.
\end{equation}
It follows that \eqref{grad flow} implies \eqref{grad constraint}.  So on a nearly parallel $G_2$-manifold, the
instanton equation on the cylinder is equivalent to the gradient flow for $W$.  In all other cases the instanton
equation on the cylinder is equivalent to the gradient flow for $W$, together with a number of constraints (see
\cite{HILP10,Xu09} for discussions of the nearly K\"ahler case).

The gradient flow structure can be seen at the level of the reduced equations.  For example, \eqref{SE gaugino 1},
\eqref{SE gaugino 2} are the gradient flow equations for
\begin{equation}
 W(\psi,\chi) = \chi^2-2\chi\psi^2+2\psi^2-1,\quad \dd s^2 = \frac{e^{2h}}{m}\dd\chi^2 + \frac{4m}{m+1}\dd\psi^2.
\end{equation}

\section{Heterotic string theory}
\label{sec:string}

The BPS equations for heterotic supergravity are
\begin{eqnarray}\label{HetSugraBPSeqtns}
\label{gravitino}
 \nabla^-\epsilon &=& 0 \\
\label{dilatino}
 (\dd\phi-H) \cdot\epsilon &=& 0 \\
\label{gaugino}
 F\cdot\epsilon &=& 0.
\end{eqnarray}
Here $H$ and $\phi$ are a 3-form and a function on a Riemannian spin manifold, $\nabla^-$ is a metric-compatible
connection with totally anti-symmetric torsion equal to $-2H$, and $F$ is the curvature of a connection on some vector
bundle.  The spinor $\epsilon$ is regarded as the generator of supersymmetries.  The three equations are known as the
gravitino equation \eqref{gravitino}, the dilatino equation \eqref{dilatino}, and the gaugino equation \eqref{gaugino}. 
In order to obtain solutions of heterotic supergravity, they must be supplemented by the Bianchi identity
\begin{equation}
\label{Bianchi}
 \dd H = -\frac{\alpha'}{4}\Tr(F\wedge F - R^+\wedge R^+).
\end{equation}
Here $\alpha'$ is the string coupling constant and $R^+$ is the curvature of the metric-compatible connection with
torsion $2H$.  Although supergravity theories exist only in specific dimensions, the equations
\eqref{gravitino}-\eqref{Bianchi} make sense in any dimension. Solutions in dimensions less than 10 can be extended to
10-dimensional ones by addition of a Minkowski space factor, and hence give rise to string theory backgrounds.

In the previous section we have constructed instantons on the cones over manifolds with real Killing spinor(s).  These
are solutions of the gaugino equation \eqref{gaugino}, with $\epsilon$ the lift of the Killing spinor(s) to the cone. 
In the present section we will extend these solutions to the full set of  equations \eqref{gravitino}-\eqref{Bianchi}. 
Our procedure generalises constructions given in \cite{Strom90,HS90,GN95} in the cases where $M=S^3$, $S^6$ and $S^7$
(with its nearly parallel $G_2$-structure).  Like in those references, we work perturbatively in $\alpha'$.  At $O(1)$,
the BPS equations are solved by the cone metric, with $H=0$ and $\phi$ constant. At $O(\alpha')$ this remains a
solution if we set the gauge field equal to the Levi-Civita connection. If, however, the gauge field equals one of the
instantons constructed above then $H$ and $\dd\phi$ are
no longer allowed to vanish, due to the coupling between $H$ and $F$ introduced by the Bianchi identity.

 A comment is in order on the equations of motion. Normally, supergravity BPS equations imply a set of second order
equations, known as the equations of motion of the theory. In heterotic supergravity, which is to be thought of as a
low-energy limit of string theory, this is only true perturbatively. The BPS equations
\eqref{HetSugraBPSeqtns} together with the Bianchi identity \eqref{Bianchi} imply the equations of motion only up to
higher order $\alpha'$-corrections, and to obtain a fully consistent theory requires including all corrections
\cite{BdR89}. In practise this can be achieved only when one has a vanishing result for higher order terms. In the following, we will
solve the
Bianchi identity perturbatively, replacing the curvature $R^+$ by the Riemannian curvature form of the cone, which
satisfies the instanton condition. This replacement can also be made in the equations of motion, and a theorem of
Ivanov tells us that the resulting BPS equations (which are unchanged) and the Bianchi identity imply the modified
equations of motion without any corrections \cite{Iv09}. Therefore, we can view our solutions either as perturbative
solutions of heterotic string theory, or as exact solutions of a heterotic supergravity which differ slightly from the
truncation of the $\alpha'$-expansion from string theory.

We work using the following metric and $G$-structure on the cone:
\begin{equation}
 \tilde g = e^{2f(\tau)} g_Z,\quad \tilde Q = e^{4f(\tau)} Q_Z,
\end{equation}
where $g_Z$ and $Q_Z$ are the metric and 4-form on the cylinder introduced in \eqref{cylinder metric}, \eqref{NKNP Q},
\eqref{SE Q}, \eqref{3S Q}.  The cone metric is of course $f=\tau$ (and $h=0$, where appropriate).  Throughout this
section, the Clifford action $\cdot$ and the contraction operator $\lrcorner$ will be assumed to be taken with respect
to this metric $\tilde g$.

\subsection{Nearly K\"ahler and nearly parallel $G_2$}

\paragraph{Dilatino equation.}  For any 1-form $v$ we have that
\begin{equation}
 v\lrcorner\tilde Q\cdot \epsilon = \kappa v\cdot\epsilon,
\end{equation}
where $\kappa=4$ in dimension 7 or 7 in dimension 8.  Thus for any $\phi=\phi(\tau)$, the dilatino equation is solved by
\begin{equation}
\label{NKNP H}
H = \frac{1}{\kappa}\dd\phi\lrcorner\tilde Q = \frac{\dot\phi}{\kappa}e^{2f}P.
\end{equation}

\paragraph{Gravitino equation.} To solve the gravitino equation, we make an ansatz for the connection $\nabla^-$ similar
to the ansatz \eqref{NKNP inst ansatz} for the gauge field:
\begin{equation}
 \nabla^- = \nabla^P + s(\tau) e^aI_a.
\end{equation}
This connection always solves equation \eqref{gravitino}, since by construction its holonomy group is contained in
$G_2$ or Spin(7), but we still need to check that its torsion is given by $H$. 
The torsion is calculated by choosing an orthonormal basis 
\begin{equation}
\tilde e^0=\exp(f) e^0, \quad\tilde e^a=\exp(f)e^a,
\end{equation}
and employing the Cartan structure equation \eqref{CSE}.  We find that $T^0=0$
and
\begin{equation}
T^a = \exp(f)\left( ( \dot f-s)e^{0a} + \frac{1-s}{\rho}P_{abc}e^{bc} \right).
\end{equation}
On the other hand, the torsion should be $T^\mu=-2\tilde e^\mu\lrcorner H$, where $H$ is the solution \eqref{NKNP H} to
the dilatino equation.  Thus we must set $s=\dot f$, and the gravitino and dilatino equations are equivalent to
\begin{eqnarray}
\label{NKNP gd 2}
 \dot f-1 &=& \frac{\rho}{\kappa}\dot\phi.
\end{eqnarray}
The general solution (satisfying the boundary condition $f-\tau \to0$ as $\tau\to\infty$) is 
\begin{equation}
\label{NKNP dilaton}
 \phi = \phi_0 + \frac{\kappa}{\rho}(f-\tau),
\end{equation}
where $\phi_0$ is the asymptotic value of $\phi$.

Notice that the torsion of $\nabla^-$ vanishes when $s=\dot f=1$.  So this connection is a torsion-free
metric-compatible connection on the cone.  This justifies our earlier claim that the connection \eqref{NKNP inst ansatz}
with $\psi=1$ is the Levi-Civita connection on the cone.

\paragraph{The Bianchi identity.}  The solution \eqref{NKNP inst ansatz}, \eqref{NKNP inst} of the instanton equation is
valid for arbitrary scale factor $f$, since the instanton equations are conformally invariant.  Thus to complete our
solution we only need to solve the Bianchi identity.  Since we are only working to leading order in $\alpha'$, the
curvature $R^+$ appearing in the Bianchi identity \eqref{Bianchi} can be replaced by the curvature $R=R^iI_i$ of the
Levi-Civita connection on the cone.  The trace appearing in the Bianchi identity will be taken using the quadratic form
that makes $I_i, I_a$ orthonormal, so that
\begin{equation}
-\Tr(F\wedge F-R^+\wedge R^+) = F^i\wedge F^i-R^i\wedge R^i + F^a\wedge F^a.
\end{equation}
These terms will be evaluated separately.

First, using the identity,
\begin{equation}
\frac{1}{4}P_{abc}P_{ade}e^{bcde} = 2\rho\, Q,
\end{equation}
we find that
\begin{eqnarray}
F^a\wedge F^a &=& \left(\dot \psi e^{0a}+\frac{\psi -\psi^2}{\rho}P_{abc}e^{bc}\right)\wedge \left(\dot \psi
e^{0a}+\frac{\psi -\psi^2}{\rho}P_{ade}e^{de}\right) \\
&=& \frac{8}{\rho}(\psi -\psi^2)^2 Q + \frac{12}{\rho}\dot \psi (\psi -\psi^2) e^0\wedge P.
\end{eqnarray}
To evaluate the remaining terms, we note that the Riemann curvature $R^a_b=R^if_{ib}^a$ satisfies the first Bianchi
identity, $\exp(\tau)e^b\wedge R^a_b=0$, and it follows that
\begin{equation}
f^i_{ab}e^{ab}\wedge R^i = 0.
\end{equation}
In addition we note that the 4-form $Q$ can be expressed as the Casimir for the structure group $K$:
\begin{equation}
 \frac{1}{4}f^i_{ab}f^i_{cd}e^{abcd} = -\frac{8}{\rho}Q.
\end{equation}
It follows from the above that
\begin{eqnarray}
 F^i\wedge F^i &=& \left(R^i + \frac{\psi^2-1}{2}f^i_{ab}e^{ab}\right) \wedge \left(R^i +
\frac{\psi^2-1}{2}f^i_{cd}e^{cd}\right) \\ &=& -\frac{8}{\rho}(\psi^2-1)^2Q + R^i\wedge R^i.
\end{eqnarray}
Thus, the Bianchi identity is
\begin{eqnarray}
\dd H &=& \frac{\alpha'}{\rho} \left( 3\dot \psi (\psi -\psi^2)e^0\wedge P + 2(-1+3\psi^2-2\psi^3)Q\right) \\
 &=& -\frac{\alpha'}{2\rho}\dd\left((1-\psi)^2(1+2\psi)P\right) .
\end{eqnarray}
Comparing with (\ref{NKNP H}), the Bianchi identity is equivalent to
\begin{equation}
 \frac{\dot\phi}{\kappa}\exp(2f) = - \frac{\alpha'}{2\rho}(1-\psi)^2(1+2\psi).
\end{equation}
Multiplying both sides of this equation by $\rho\exp(-2\tau)$ and employing equations (\ref{NKNP inst}) and
(\ref{NKNP gd 2}) gives the equation,
\begin{equation}
 \frac{1}{\rho}(\dot f-1)\exp(2(f-\tau)) = \frac{\alpha'}{2\rho}\exp(-2\tau)( -1+\psi^2 - \psi\dot \psi),
\end{equation}
which can in fact be integrated exactly:
\begin{equation}
 e^{2f} = e^{2\tau} + \frac{\alpha'}{2}(1-\psi^2).
\end{equation}
Together with \eqref{NKNP inst}, \eqref{NKNP dilaton}, this gives a solution of the gaugino, gravitino and dilatino
equations and the Bianchi identity.  The constant $\phi_0$ is the background value of the dilaton field and $\tau_0$ is
a parameter controlling the instanton size.  In the cases $M=S^6,S^7$ this reproduces solutions constructed in
\cite{HS90,GN95}.

\subsection{Sasaki-Einstein}

\paragraph{Dilatino equation.}  For any 1-form $v$ we have that
\begin{equation}
 v\lrcorner\tilde Q\cdot \epsilon = m v\cdot\epsilon.
\end{equation}
Thus for any $\phi=\phi(\tau)$, the dilatino equation is solved by
\begin{equation}
\label{SE H}
H = \frac{1}{m}\dd\phi\lrcorner\tilde Q = \frac{\dot\phi}{m}e^{2(f+h)}P.
\end{equation}

\paragraph{Gravitino equation.} To solve the gravitino equation, we make an ansatz for the connection $\nabla^-$ similar
to \eqref{SE inst ansatz}
\begin{equation}
 \nabla^- = \nabla^P + t(\tau) e^1I_1 + s(\tau) e^aI_a.
\end{equation}
This has holonomy $\mbox{SU}(m+1)$, so solves \eqref{gravitino}.  To calculate the torsion of $\nabla^-$, we choose an
orthonormal basis 
\begin{equation}
\tilde e^0=\exp(f) e^0, \quad\tilde e^a=\exp(f+h)e^a,\quad\tilde e^1=\exp(f)e^1,
\end{equation}
and employ the Cartan structure equation \eqref{CSE}.  We find that $T^0=0$ and
\begin{equation}
\begin{aligned}
T^1 &= e^f\left( ( \dot f - t)e^{01} + (1-e^hs)P_{1ab}e^{ab} \right) \\
T^a &= e^{f+h}\left\{ ( \dot f + \dot h - e^{-h}s)e^{0a} + ((1-e^{-h}s)+(1-t)/m)P_{ab1}e^{b1} \right\}.
\end{aligned}
\end{equation}
On the other hand, the torsion should be $T^\mu=-2\tilde e^\mu\lrcorner H$, with $H$ given by \eqref{SE H}.  Thus we set
$t=\dot f$, $s=e^h(\dot f + \dot h)$, so that the gravitino and dilatino equations are equivalent to
\begin{eqnarray}
 \label{SE gd 3}
 \frac{2}{m}\dot\phi &=& \frac{m+1}{m}(\dot f-1)+\dot h \\
 \label{SE gd 4}
 \frac{m-1}{m}\dot f + \dot h &=& 2e^{-2h}-\frac{m+1}{m}.
\end{eqnarray}
Equation \eqref{SE gd 3} can be integrated to give $\phi$ in terms of $f$ and $h$:
\begin{equation}
 \phi = \phi_0 + \frac{m+1}{2}(f-\tau) + \frac{m}{2}h.
\end{equation}

Notice that the torsion of $\nabla^-$ vanishes when $s=t=\dot f=1$, $h=0$.  So this connection is a torsion-free
metric-compatible connection on the cone.  This justifies our earlier claim that the connection \eqref{SE inst ansatz}
with $\chi=\psi=1$ is the Levi-Civita connection on the cone.

\paragraph{The Bianchi identity.}  The trace appearing in the Bianchi identity will be normalised so that the $I_a$ are orthonormal.  This convention
implies that $-\Tr(I_1^2)=(m+1)/2m$.  The $I_i$ will be taken to be orthonormal also.  Thus,
\begin{equation}
-\Tr(F\wedge F-R^+\wedge R^+) = F^i\wedge F^i-R^i\wedge R^i + F^a\wedge F^a + \frac{m+1}{2m}F^1\wedge F^1.
\end{equation}
Here as above $R^+$ has been replaced by the curvature $R=R^iI_i$ of the Levi-Civita connection on the cone, since we
are working only to leading order in $\alpha'$.  These terms will be evaluated separately.

First, using the identity,
\begin{equation}
P_{1ab}P_{1cd}e^{abcd} = 8 Q,
\end{equation}
we find that
\begin{eqnarray}
F^a\wedge F^a &=& 4\frac{m+1}{m}\dot\psi\psi(1-\chi)e^0\wedge P \\
\frac{m+1}{2m} F^1\wedge F^1 &=& 2\frac{m+1}{m}\dot\chi(\chi-\psi^2)e^0\wedge P + 4\frac{m+1}{m}(\chi-\psi^2)^2Q.
\end{eqnarray}
From the first Bianchi identity for $R$ it follows that
\begin{eqnarray}
 F^i\wedge F^i &=& -4\frac{m+1}{m}(\psi^2-1)^2Q + R^i\wedge R^i.
\end{eqnarray}
Thus, the Bianchi identity is
\begin{eqnarray}
\nonumber
\dd H &=& \frac{\alpha'(m+1)}{4m} \left( 4\dot\psi\psi(1-\chi)+2\dot\chi(\chi-\psi^2) \right) e^0\wedge P \\
&& + \frac{\alpha'(m+1)}{m} \left( \chi^2-2\chi\psi^2+2\psi^2-1 \right)Q \\
&=& \frac{\alpha'(m+1)}{4m} \dd \left((\chi^2-2\chi\psi^2+2\psi^2-1)P \right).
\end{eqnarray}
Comparing with (\ref{SE H}), (\ref{SE gd 3}), (\ref{SE gd 4}) the Bianchi identity reduces to
\begin{equation}
\label{SE bianchi}
 (\dot f + \dot h - e^{-2h})e^{2(f+h)} = \frac{\alpha'(m+1)}{4m} \left(\chi^2-2\chi\psi^2+2\psi^2-1 \right).
\end{equation}

\begin{figure}[!ht]
 \psfrag{p0}{$\psi_0$}
 \psfrag{c0}{$\chi_0$}
 \psfrag{p1}{$\psi_1$}
 \psfrag{c1}{$\chi_1$}
 \psfrag{h1}{$h_1$}
 \psfrag{f1}{$f_1$}
 \psfrag{tau}{$\tau$}
 \epsfig{ file = 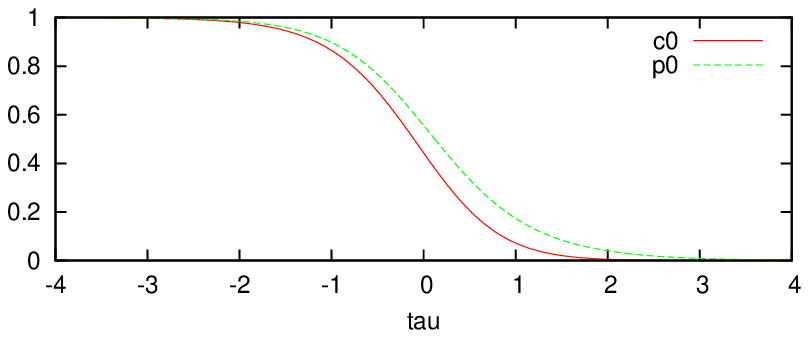, scale = 1.3 }
 \epsfig{ file = 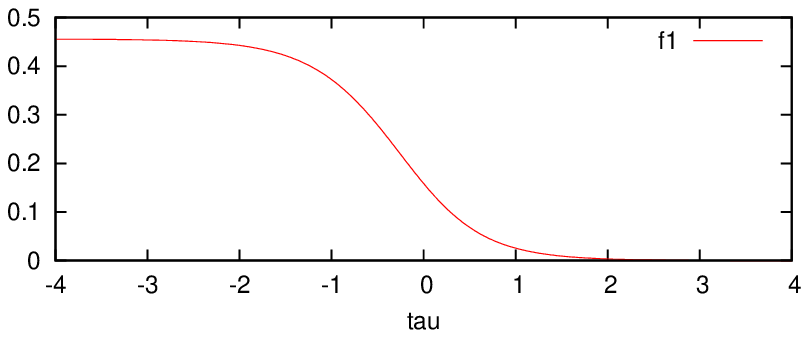, scale = 1.3 }
 \epsfig{ file = 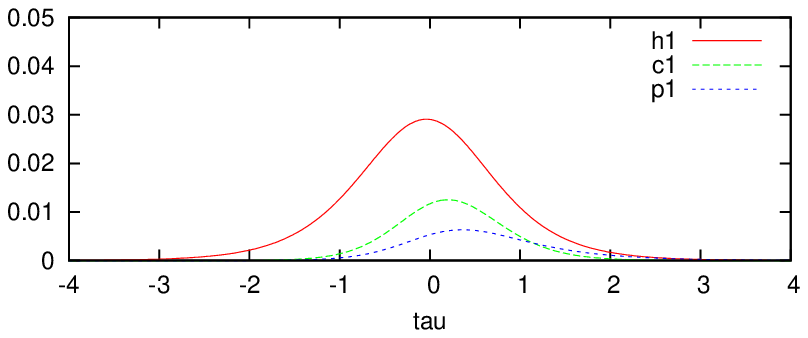, scale = 1.3 }
 \caption{Sample solution of the Sasaki-Einstein supergravity equations with $m=2$.}
 \label{fig2}
\end{figure}

We have reduced the heterotic supergravity equations to 4 equations \eqref{SE gaugino 1}, \eqref{SE gaugino 2}, \eqref{SE gd 4}, \eqref{SE bianchi}.  In the case $m=1$ these are solved exactly \cite{Strom90} by \eqref{SE instSol}, $h=0$, and
\begin{equation}
  e^{2f} = e^{2\tau} + \frac{\alpha'}{2}(1-\chi^2).
\end{equation}

For $m>1$ solutions may be obtained only numerically.  We assume that the solutions can be expanded in $\alpha'$:
\begin{equation}
 \begin{aligned}
  \chi &= \chi_0 + \alpha'\chi_1, & f &= f_0 + \alpha'f_1, \\
  \psi &= \psi_0 + \alpha'\psi_1, & h &= h_0 + \alpha'h_1.
 \end{aligned}
\end{equation}
The functions $\chi_0,\psi_0,f_0,h_0$ are solutions at $O(1)$ in $\alpha'$.  The unique $O(1)$ solution of \eqref{SE gd 4}, \eqref{SE bianchi} for which $h_0$ does not blow up at $\tau=-\infty$ is $f_0=\tau$, $h_0=0$.  Then $\psi_0,\chi_0$ must solve \eqref{SE gaugino 1}, \eqref{SE gaugino 2} with $h=0$.  As discussed in section \ref{sec:cone}, there is a 1-parameter family of solutions which do not blow up, with a translational parameter $\tau_0$.  At $O(\alpha')$, equations \eqref{SE gaugino 1}, \eqref{SE gaugino 2}, \eqref{SE gd 4}, \eqref{SE bianchi} reduce to the following differential equations for $\chi_1,\psi_1,f_1,h_1$:
\begin{eqnarray}
\label{SE pert 1}
 \dot h_1 &=& -2(m+1)h_1 - \frac{m^2-1}{4m}e^{-2\tau}\left(\chi_0^2-2\chi_0\psi_0^2+2\psi_0^2-1 \right) \\
 \label{SE pert 2}
 \dot f_1 &=& 2mh_1 + \frac{m+1}{4}e^{-2\tau}\left(\chi_0^2-2\chi_0\psi_0^2+2\psi_0^2-1 \right) \\
 \label{SE pert 3}
 \dot \chi_1 &=& -4mh_1(\psi_0^2-\chi_0) + 4m\psi_0\psi_1 - 2m\chi_1 \\
 \label{SE pert 4}
 \dot \psi_1 &=& \frac{m+1}{m}(\chi_0-1)\psi_1 + \frac{m+1}{m}\psi_0\chi_1.
\end{eqnarray}
We assume that solutions of these equations exist for all $\tau$.  Solutions $h_1,\chi_1,\psi_1$ of \eqref{SE pert 1}, \eqref{SE pert 3}, \eqref{SE pert 4} may blow up as $\tau\to-\infty$, and for each $\tau_0$ there is a unique solution which does not.  Then equation \eqref{SE pert 2} has a unique solution satisfying $f_1\to0$ as $\tau\to\infty$.  So, the supergravity equations have a 1-parameter family of solutions to $O(\alpha')$.  These solutions have the following asymptotics:
\begin{equation}
 \begin{aligned}
  1-\chi_0,1-\psi_0,h_1,\dot f_1 &\sim e^{2\tau},& \chi_1,\psi_1 &\sim e^{4\tau} && \mbox{as } \tau\to-\infty; \\
  \psi_0,\psi_1 \sim e^{-\frac{m+1}{m}\tau},\quad  \chi_0,\chi_1 &\sim e^{-2\frac{m+1}{m}\tau}, & h_1,f_1&\sim e^{-2\tau} && \mbox{as } \tau\to\infty.
 \end{aligned}
\end{equation}

We have constructed numerical solutions using a Runge-Kutta algorithm.  The boundary condition $h_1=\chi_1=\psi_1=0$ was imposed at a large (but finite) negative value of $\tau$.  We have checked that these numerical solutions have the correct asymptotics as $\tau\to-\infty$, and our algorithm reproduces the exact solutions when $m=1$.  A sample solution with $m=2$ is displayed in figure \ref{fig2}.  The asymptotics at $\tau=-\infty$ guarantee that when $M=S^{2m+1}$, our supergravity solutions extend over the apex of the cone.  Thus we obtain solutions of the supergravity equations in $\RR^{2m+2}$.

\subsection{3-Sasakian}  

\paragraph{Gravitino equation.}  Our ansatz for $\nabla^-$ is similar to \eqref{3S inst ansatz}:
\begin{equation}
 \nabla^- = \nabla^P + t(\tau) e^\alpha I_\alpha + s(\tau) e^ a I_a.
\end{equation}
This solves the gravitino equation \eqref{gravitino}, where $\epsilon$ are given by the $m+2$ Killing spinors
$\epsilon_A$.  Introducing the orthonormal basis
\begin{equation}
 \tilde e^ 0 = \exp(f)\dd\tau,\quad \tilde e^\alpha = \exp(f)e^\alpha ,\quad \tilde e^a = \exp(f+h) e^a
\end{equation} 
and using the Cartan structure equation \eqref{CSE} we find that $T^0=0$ and
\begin{equation}
 \begin{aligned}{}
  T^\alpha &= e^f \Big\{(\dot f-t)e^{0\alpha} +2(1-e^h s) \omega^\alpha + (1-t) \varepsilon_{\alpha\beta\gamma}
e^{\beta\gamma} \Big\} \\
  T^a &= e^{f+h} \Big\{ (\dot f+\dot h -e^{-h}s) e^{0a} +(1-e^{-h}s) \omega^\alpha_{ab} e^{b\alpha} \Big\}.
 \end{aligned}
\end{equation} 
Skew-symmetry of the torsion requires it to be of the form $T^\mu = -2\tilde e^\mu \lrcorner H$ for some 3-form $H$. 
This means in particular that we must set $t=\dot f$ and $s=e^h(\dot f+\dot h)$.  Then the torsion will be
skew-symmetric if and only if
\begin{equation}\label{3S-connTorsion}
 \dot f+\dot h = 2e^{-2h}-1.
\end{equation} 
Assuming that \eqref{3S-connTorsion} holds,  the 3-form $H$ is given by
\begin{equation}\label{3S H}
 H= e^{2f }(\dot f -1)\eta^{123} + \frac{1}{2}e^{2(f+h)}\left(\dot f+\dot h-1\right)\eta^\alpha \wedge
\omega^\alpha .
\end{equation}
Notice that when $s=t=\dot f=e^h=1$ the torsion vanishes.  This justifies our earlier claim that the connection \eqref{3S
inst ansatz} with $\psi=\chi=1$ is the Levi-Civita connection on the cone.

\paragraph{Dilatino equation.}  To solve the dilatino equation we make use of the following identities
\begin{equation}
  \eta^{123} \cdot \epsilon = e^{-2f} \dd\tau \cdot \epsilon, \qquad \eta^\alpha\wedge \omega^\alpha \cdot \epsilon = e^{-2(f+h)} 2m\, \dd\tau\cdot \epsilon.
\end{equation} 
Thus, if $\phi$ is a function of $\tau$ and $H$ is given by \eqref{3S H}, the dilatino equation \eqref{dilatino} is
equivalent to
\begin{equation}
 \dot\phi = (m+1)(\dot f-1) + m\dot h.
\end{equation}
Clearly, this is solved by
\begin{equation}
\label{3SDilatinoSoln}
 \phi = \phi_0 + (m+1)(f-\tau) + m h,
\end{equation}
where the integration constant $\phi_0$ may be interpreted as the background value of the dilaton field.

\paragraph{The Bianchi identity.}  The instanton that we constructed in the previous section solves the gaugino equation
on the cone for all possible choices of the functions $f,h$.  Thus it remains to solve the Bianchi identity
\eqref{Bianchi}.  Working to leading order in $\alpha'$, we shall replace $R^+$ by $R=R^iI_i$, the Riemann curvature
form of the cone.  We have
\begin{equation}
 F^i = R^i + \frac 12 (\psi^2-1) f^i_{ab} e^{ab}.
\end{equation} 
Now we can calculate the terms occurring in the Bianchi identity (without at this point assuming that $\psi,\chi$
solve the instanton equation):
\begin{equation}
 \begin{aligned}{}
  \frac12 F^\alpha \wedge F^\alpha =&\; 2\dot\chi(\chi-\psi^2)e^0\wedge \eta^\alpha\wedge\omega^\alpha +
6\dot\chi(\chi-\chi^2)e^0\wedge\eta^{123} \\
  &\; + 2(\chi-\psi^2)(\chi-\chi^2)\epsilon_{\alpha\beta\gamma}\eta^{\alpha\beta}\wedge\omega^\gamma +
2(\chi-\psi^2)^2\omega^\alpha\wedge\omega^\alpha \\ 
  F^a \wedge F^ a =&\; 4\dot\psi\psi(1-\chi)e^0\wedge \eta^\alpha\wedge\omega^\alpha
-2\psi^2(1-\chi)^2\epsilon_{\alpha\beta\gamma}\eta^{\alpha\beta}\wedge\omega^\gamma \\
  F^i \wedge F^i =&\; R^i \wedge R^i -2 (\psi^2-1)^2 \omega^\alpha\wedge\omega^\alpha.
 \end{aligned}
\end{equation} 
Here we have used the first Bianchi identity $R^i f_{ia}^b \tilde e^a=0$, as well as the following formula for the
Casimir 4-form
\begin{equation}
 f^i_{ab} f^i_{cd} e^{abcd} = -8 \omega^\alpha \wedge \omega^\alpha.
\end{equation} 
We assume that the trace has been normalised so that $I_a, I_i$ are orthonormal, then we must have that $-\Tr(I_\alpha
I_\beta) = 1/2\delta_{\alpha\beta}$. Thus the Bianchi identity is
\begin{eqnarray}
 \dd H  &=& \frac{\alpha'}{4}\left(F^i\wedge F^i - R^i \wedge R^ i + F^a \wedge F^a  + \frac 12  F^\alpha \wedge F^\alpha\right) \\
 &=& \frac{\alpha'}{4}\dd\left[ (\chi-1)\left\{(1+\chi-2\chi^2)\eta^{123} + (1+\chi-2\psi^2)\eta^\alpha\wedge\omega^\alpha\right\}\right].
\end{eqnarray}
Now we assume that the gauge field is an instanton, so that in particular $\chi=\psi^2$ (cf.\ \eqref{3SInstNecCond1}). 
The Bianchi identity is solved by
\begin{equation}
 H = - \frac{\alpha'}{4} \left[ (1+2\chi)(1-\chi)^2\eta^{123} + (1-\chi)^2\eta^\alpha\wedge\omega^\alpha \right].
\end{equation}
Comparing with our earlier solution \eqref{3S H} of the gravitino equation, we see that the Bianchi identity, gravitino
equation, and gaugino equation are equivalent to
\begin{eqnarray}
\label{3SBianchi1}
 e^{2f }\left(\dot f -1\right) &=& - \frac{\alpha'}{4} (1+2\chi)(1-\chi)^2, \\
\label{3SBianchi2}
 e^{2(f+h)}\left(\dot f+\dot h-1\right) &=& - \frac{\alpha'}{2}(1-\chi)^2,
\end{eqnarray}
together with equation \eqref{3S-connTorsion}, where $\chi$ is given by the solution \eqref{3SInstSoln1} to the
differential equation \eqref{3SInstNecCond2}.  Note that once again we have more equations than unknowns, so naively one
would not expect this system to have any solutions.  In spite of this, an analytic solution can be found: it is
\begin{eqnarray}
\label{3SBianchiSoln1}
 e^{2f} &=& e^{2\tau} + \frac{\alpha'}{4}(1-\chi^2) \\
\label{3SBianchiSoln2}
 e^{2(f+h)} &=& e^{2\tau} + \frac{\alpha'}{2}(1-\chi).
\end{eqnarray}
Thus we have obtained a 1-parameter family of solutions of the gaugino, gravitino and dilatino equations and the Bianchi identity, with the
functions $\chi$, $\psi$, $f$, $h$ and $\phi$ given in equations \eqref{3SInstSoln1}, \eqref{3SInstSoln2}, \eqref{3SBianchiSoln1},
\eqref{3SBianchiSoln2} and \eqref{3SDilatinoSoln}.

 In the limits $\tau \rightarrow \pm \infty$ we get
 \begin{equation}
  \begin{aligned}{}
    \tau \rightarrow -\infty:\qquad & h\rightarrow 0,\quad \chi,\psi \rightarrow 1,\quad e^{2f} \rightarrow
e^{2\tau}\left(1
+\frac {\alpha'}2 \right) \\
     \tau \rightarrow +\infty:\qquad & h\rightarrow 0,\quad \chi,\psi \rightarrow 0,\quad e^{2f} \rightarrow
e^{2\tau} .
  \end{aligned}
 \end{equation} 
 The limiting behaviour is very similar to the nearly K\"ahler and nearly parallel $G_2$ cases. In
particular, the metric equals the Ricci-flat cone metric in both limits, and the instanton approaches the
canonical connection for $\tau \rightarrow \infty$ and the Levi-Civita connection on the cone for $\tau \rightarrow
-\infty$.  In the particular case $M=S^{4m+3}$ the solution extends over the apex of the cone: thus the quaternionic instanton of
\cite{CGK85,BIL08} lifts to heterotic supergravity on $\RR^{4(m+1)}$.

\section*{Acknowledgements}
 
We are grateful to Alexander Popov for carefully reading the manuscript.  This work was done within the framework of the project supported by the DFG under the grant 436 RUS 113/995.  D.~H.\ is supported by EPSRC through the grant EP/G038775/1.

\end{document}